\documentclass[twocolumn,superscriptaddress,showpacs,preprintnumbers,amsmath
,amssymb,aps,prd]{revtex4}
\usepackage{graphicx}
\usepackage{slashbox}
\usepackage{bm}
\usepackage{color}
\graphicspath{{Figures/}}

\def\bh#1{black hole#1
  (BH#1)\gdef\bh{BH}}
\def\bbh#1{binary black hole#1
  (BBH#1)\gdef\bbh{BBH}}
\def\gw#1{gravitational wave#1
  (GW#1)\gdef\gw{GW}}
\def\nr#1{numerical relativity#1
  (NR#1)\gdef\nr{NR}}
\def\snr#1{signal-to-noise-ratio#1
  (SNR#1)\gdef\snr{SNR}}

\makeatletter
\let\protect\relax
{\catcode`\|=\active
  \xdef\InnerProduct{\protect\expandafter\noexpand\csname InnerProduct
\endcsname}
  \expandafter\gdef\csname InnerProduct \endcsname#1{%
    \begingroup
    \ifx\SavedDoubleVert\relax
    \let\SavedDoubleVert\|\let\|\IpDoubleVert
    \fi
    \mathcode`\|32768\let|\IPVert
    \left({#1}\right)
    \endgroup
  }
}
\def\IPVert{\@ifnextchar|{\|\@gobble}
     {\egroup\,\mid@vertical\,\bgroup}}
\def\IPDoubleVert{\egroup\,\mid@dblvertical\,\bgroup}
\let\SavedDoubleVert\relax
\def\midvert{\egroup\mid\bgroup}
\def\SetVert{\@ifnextchar|{\|\@gobble}
    {\egroup\;\mid@vertical\;\bgroup}}
\def\SetDoubleVert{\egroup\;\mid@dblvertical\;\bgroup}
\def\mid@vertical{\mskip1mu\vrule\mskip1mu}
\def\mid@dblvertical{\mskip1mu\vrule\mskip2.5mu\vrule\mskip1mu}
\makeatother

\usepackage{braket}
\newcommand{\Overlap}{\Braket}

\begin{document}

\title{Impact of Higher-order Modes on the Detection of Binary Black Hole Coalescences}
\author{Larne Pekowsky}
\affiliation{Center for Relativistic Astrophysics and
School of Physics\\
Georgia Institute of Technology, Atlanta, GA 30332}
\author{James Healy}
\affiliation{Center for Relativistic Astrophysics and
School of Physics\\
Georgia Institute of Technology, Atlanta, GA 30332}
\author{Deirdre Shoemaker}
\affiliation{Center for Relativistic Astrophysics and
School of Physics\\
Georgia Institute of Technology, Atlanta, GA 30332}
\author{Pablo Laguna}
\affiliation{Center for Relativistic Astrophysics and
School of Physics\\
Georgia Institute of Technology, Atlanta, GA 30332}

\begin{abstract} 
The inspiral and merger of black-hole binary systems are a promising
source of gravitational waves for the array of advanced
interferometric ground-based gravitational-wave detectors currently
being commissioned.  The most effective method to look for a signal
with a well understood waveform, such as the binary black hole signal,
is matched filtering against a library of model waveforms.  While
current model waveforms are comprised solely of the dominant radiation
mode, the quadrupole mode,  it is known that there can be significant
power in the higher-order modes for a broad range of physically
relevant source parameters during the merger of the black holes.  The
binary black hole waveforms produced by numerical relativity are
accurate through late inspiral, merger, and ringdown and include the
higher-order modes.   The available numerical-relativity waveforms
span an increasing portion of the physical parameter space of unequal
mass, spin and precession.  In this paper,  we investigate the degree
to which gravitational-wave searches could be improved by the
inclusion of higher modes in the model waveforms, for signals with a
variety of initial mass ratios and generic spins.  Our investigation
studies how well the quadrupole-only waveform model matches the signal
as a function of the inclination and orientation of the source and how
the modes contribute to the distance reach into the Universe of
Advanced LIGO for a fixed set of internal source parameters.  The
mismatch between signals and quadrupole-only waveform can be large,
dropping below 0.97 for up to 65\% of the source-sky for the
non-precessing cases we studied, and over a larger area in one
precessing case.  There is a corresponding 30\% increase in detection
volume that could be achieved by adding higher modes to the search;
however, this is mitigated by the fact that the mismatch is largest
for signals which radiate the least energy and to which the  search is
therefore least sensitive.  Likewise, the mismatch is largest in
directions from the source along which the least energy is radiated.
\end{abstract}


\maketitle

\section{Introduction}

The merger of a \bbh{} system has long been considered a
strong source of gravitational waves for ground and space based
gravitational wave observatories.  These mergers are characterized by
15 parameters,  9 intrinsic to the black-hole systems (2 black-hole
masses, 2 spin vectors and eccentricity) and 6 extrinsic to the source (binary orientation vector, sky position and distance).  The LIGO
and Virgo detectors have recently completed a joint run during which
inspiral horizon distances exceeded 40 Mpc~\cite{LIGO:2012aa} and
new upper limits have been placed on the rates of such
events~\cite{PhysRevD.85.082002}.  These observatories are currently
being upgraded and when the new design sensitivities are achieved they
will have ranges up to ten times greater and hence volumes 1000 times
greater.  By the end of this decade LIGO and Virgo, along with GEO,
will be joined by KAGRA in Japan and possibly the proposed LIGO India,
greatly increasing not only the range of the global network but also
the ability to recover information about the
sources~\cite{Fairhurst:2012tf}. 

When the theoretical model of the gravitational waveform is well
understood,  the most effective method to search  and recover  a
gravitational wave signal  is matched filtering against a library of
model waveforms called a {\it template bank} \cite{Thorne1987}. 
The ability of such a templated search to detect signals is dependent
on four factors:
\begin{itemize}
\item The frequency-dependent sensitivity of the detector.  Throughout
this paper we use the targeted aLIGO \emph{zero-detuned,
high-power}~\cite{T0900288} sensitivity curve.

\item The direction-dependent sensitivity of the detector.  This is
a fixed property of interferometric instruments and the orientation
on the Earth's surface.  Any one detector will have blind spots, one
motivation for constructing a network of detectors is to provide more
complete coverage of the sky.  We will not consider multi-detector
searches in this paper.

\item The total energy radiated by the source from the time it enters
the sensitive band of the detectors.  This provides an upper limit on
the ability to detect different signals; a source that radiates less
energy will be visible out to a smaller distance than one that
radiates more energy, all other factors being equal.

\item The ability of the templates to extract signal power from the
background noise.
\end{itemize}
In this paper we will be concerned with the last two points.

For the
\bbh{} systems potentially observable by ground-based detectors,
astrophysical processes place few constraints on the intrinsic
physical parameters that characterize the emission of radiation from
these cataclysmic events, thus placing the burden on source models to
cover nearly the full compliment of physical parameters.  Rigorous
requirements from matched filtering place  an additional burden on the
source models.  In order for the model waveforms to match  potential
signals to within a given tolerance, we need not only enough waveforms
to cover the parameter space but also each waveform must  represent
nature effectually  enough to
ensure the signal does not fall through cracks in the template bank and faithfully enough to recover the source parameters.

\begin{figure}[h]
\hbox{
\includegraphics[width=.52\linewidth]{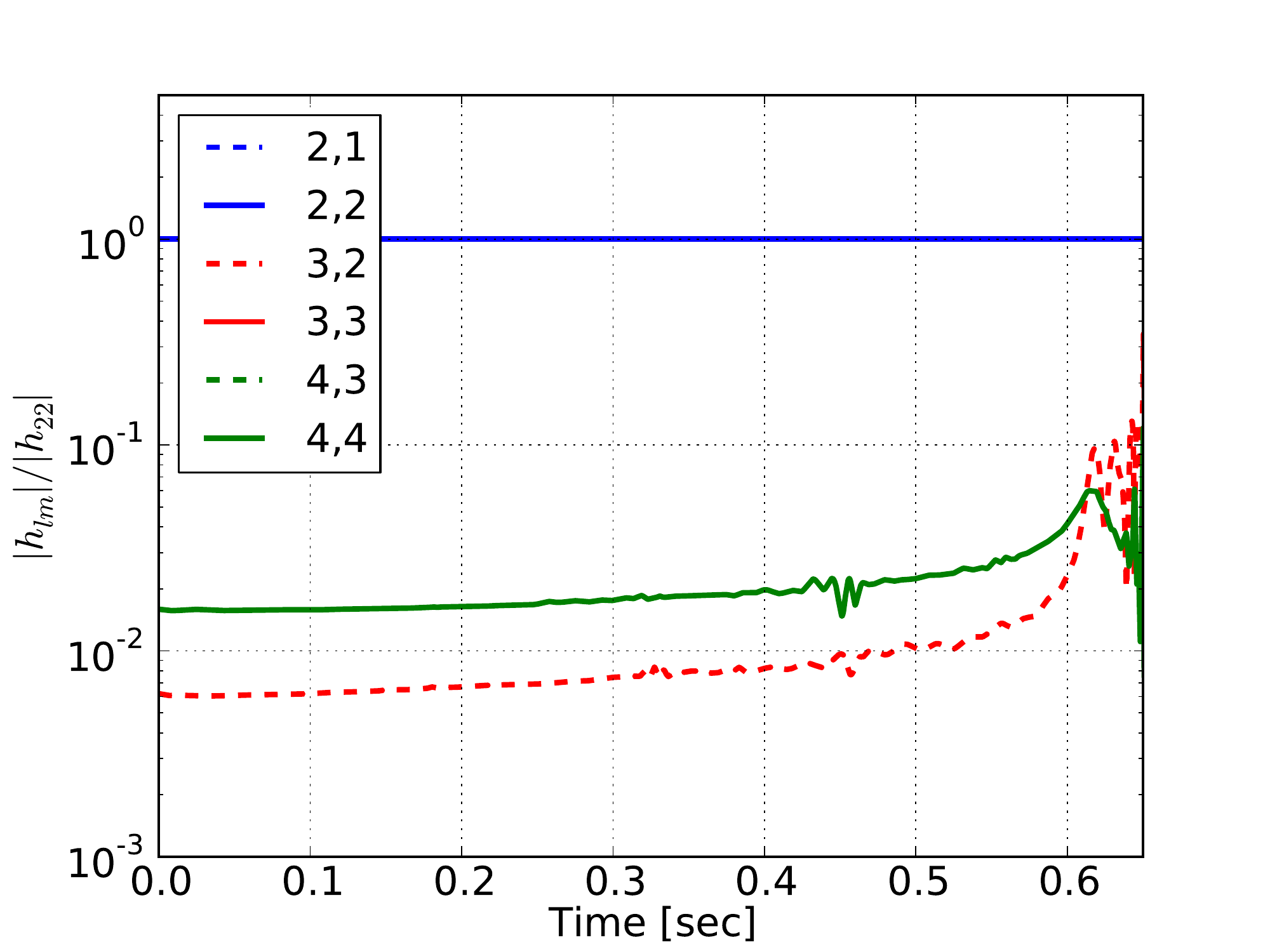}
\includegraphics[width=.52\linewidth]{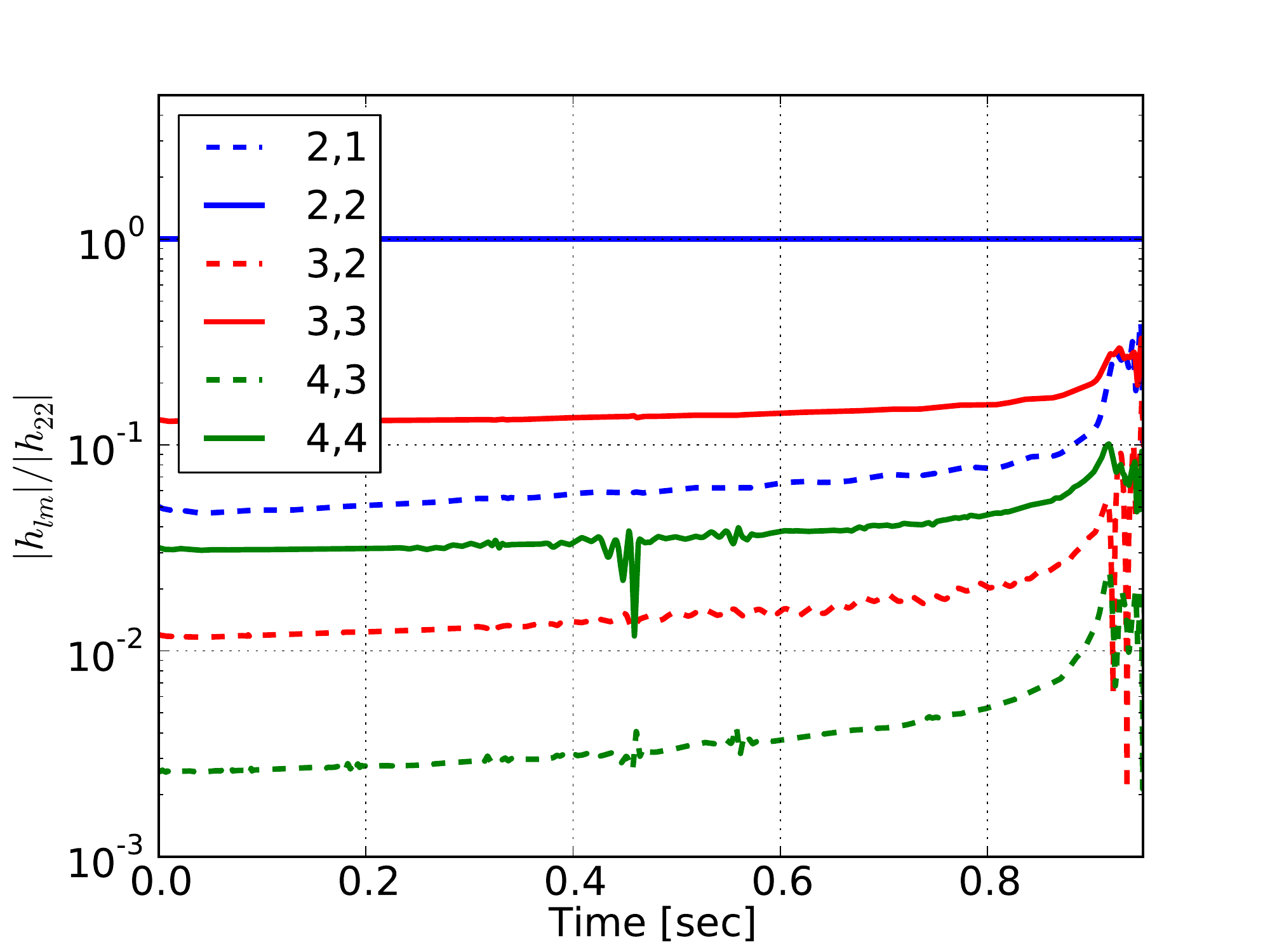}
}
\caption{Relative amplitude of higher modes for non-spinning {\bf
Left}: $q=1$ and {\bf Right}: $q=4$ systems.  For the $q=1$ system the
(4,4) and (3,2) modes are about two orders of magnitude smaller than
the (2,2).  All others are less than $10^{-3}$.  For the $q=4$ the
(3,3) mode is within a factor of 10 of the dominant (2,2) mode, and several
other modes are within another factor of ten.
}
\label{fig:modeAmplitudes}
\end{figure}

One source of mismatch with nature is the truncation of the spherical  harmonic series in which we have decomposed the model waveform.
Current template waveforms are only of the dominant, quadrupole mode, although we know that generic signals will have many excited harmonics present when detected.   Fig.(\ref{fig:modeAmplitudes}) shows the ratio of several non-dominant modes to the dominant mode for two non-spinning systems, note that for the system where the masses of the component holes are not equal the
next-to-leading mode is within an order of magnitude of the quadrupole
mode, suggesting that accounting for additional modes may be important
for detection, especially as the mass-ratio strongly deviates from one and generic spins are explored.

This paper builds on previous work by ourselves and other authors. 
In~\cite{Shoemaker:2008pe,Vaishnav:2007nm}, we conducted a preliminary study on higher modes for spinning, equal-mass systems comparing numerical relativity waveforms containing the largest five harmonics to an equal-mass non-spinning system of just the dominant mode.   We found that for low spins, the non-spinning dominant mode was an effective model waveform. McWilliams et al \cite{McWilliams:2010eq} found that  over a range of the source orientations, the equal-mass waveform was effective at detecting moderate mass ratios over source orientations.   Brown et al\cite{Brown2012} is exploring the value added of higher modes in EOBNR models of unequal-mass waveforms.  
 
In this paper we investigate the degree to which inclusion of
additional terms of the spherical harmonic series to template
waveforms could improve matched-filter based searches.  We use \nr{}
waveforms as both signal and template, and we consider both unequal
masses and some generic spins generated by the \textsc{Maya} code.  We
study how well the quadrupole-only model waveform matches the signal
as a function of the inclination and orientation of the source and
determine how the volume reach of advanced LIGO depends on the
inclusion/exclusion of non-dominant harmonics in the model waveforms.
We concentrate on system masses greater than $100 M_\odot$ to give the
NR portion of the waveform prominence and negating the need for
post-Newtonian information.  Our findings show that for non-precessing
signals up to 65\% of source orientations can be missed when using
only the quadrupole mode, implying a 30\% gain in detection volume
which could be achieved by including higher modes.  For our most
precessing case when using the quadrupole mode only the loss of source
orientations is 83\% and the potential gain in volume over which such
systems could be detected is again 30\%.  These potential gains in
volume are mitigated by the fact that the mismatch is largest for
signals which radiate the least energy and to which, therefore,  the
search is therefore least sensitive. Likewise, the mismatch is largest
in directions from the source along which the least energy is
radiated.  Finally, we do a preliminary investigation into how the
series truncation might impact parameter estimation by exploring a
potential degeneracy between mass and inclination for full waveforms
in the last section of this paper.

We proceed as follows: in $\S~\ref{sec:matchedfilter}$ we introduce
our methodology for matched filtering, and in
$\S~\ref{sec:bbhWaveforms}$ the \nr{} waveforms used in all of our
studies. In $\S$~\ref{sec:preliminaryInvestigations} we consider
various aspects of the overlaps between the dominant mode and the higher modes.  In $\S$~\ref{sec:results} we  examine the
volume of the universe accessible to advanced detectors using quadrupole-only waveforms and hypothetical ideal
waveforms containing most of the modes, for several cases.   We conclude in
$\S$~\ref{sec:conclusions} that the smallest overlaps are obtained for
systems and source orientations which radiate the least total power,
and hence have the smallest accessible volumes even when an ideal waveform is used.  In this section we also present a first look at the
implications of higher modes for parameter estimation.

\begin{figure}[h]
{\centering
\includegraphics[width=0.44\linewidth]{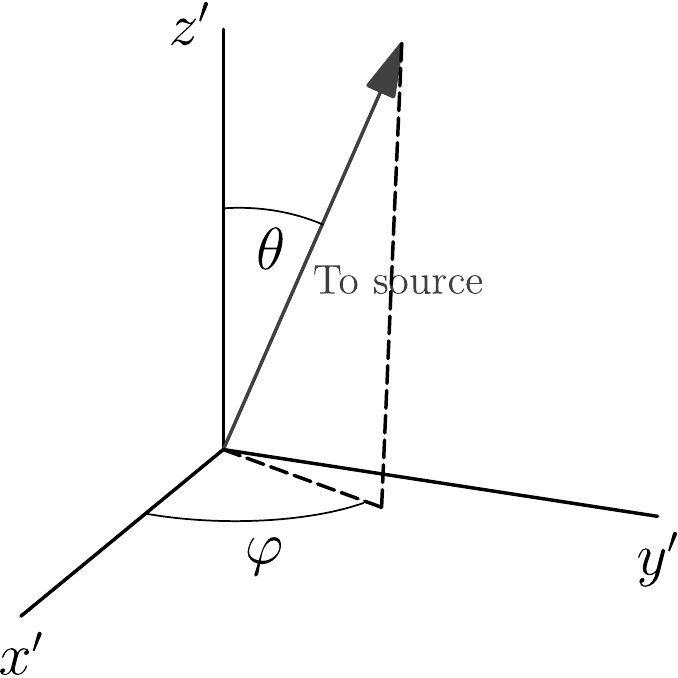}
\includegraphics[width=0.44\linewidth]{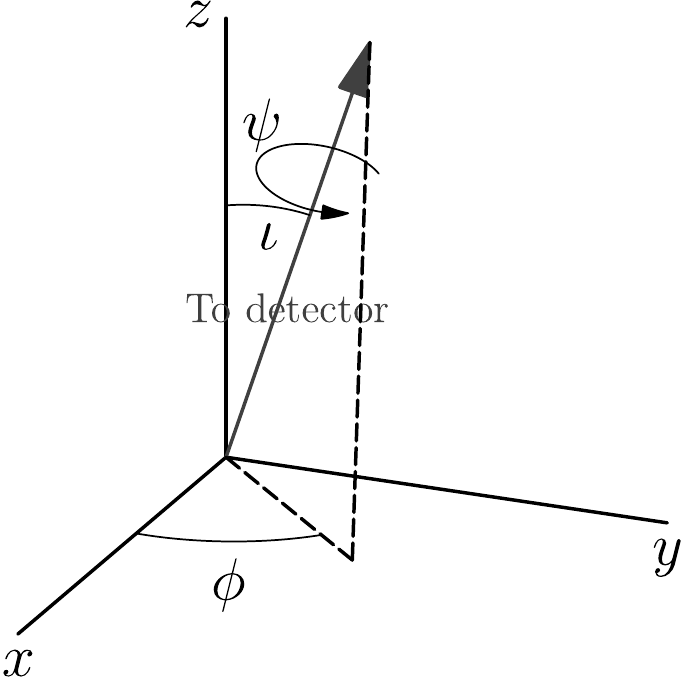}
}
\caption{
Definition of angles used in this paper.  {\bf Left}: the angles used
at the detector, looking at the source.  Although $\psi$ refers to a
rotation of the plane containing the source, we associate it with the
detector because it enters the analysis though the antenna pattern.
{\bf Right}: the angles used at the source, looking towards the
detector.  These are the angles in which the spherical harmonics are
written.}
\label{fig:angles}
\end{figure}

\emph{Conventions:} Throughout this paper we adopt the following
conventions.  We denote the Fourier transform of a function $g(t)$
with a tilde, as $\tilde{g}(f)$.  We characterize the mass ratio of a
\bbh{} system by $q=m_1/m_2$ with $m_1 \ge m_2$.  The relation of the
source to the detector is specified by five angles.  Two ($\iota,
\phi$) place the detector in coordinates centered at the source, it is
these angles in which the decomposition into spherical harmonics is
performed.  Two ($\theta, \varphi$) place the source in the sky of the
detector.  The final angle, $\psi$, determines the relative rotation
between these two coordinate systems, we associate $\psi$ with the
source because in what follows we will treat it similarly to $\iota$
and $\phi$.  We define these angles in fig.(\ref{fig:angles}).  The
final parameter connecting the source and detector is the distance
between them, we will be concerned with the maximum distance at which
the source can be detected and will determine this value in what
follows.

\section{Matched-filter searches for Gravitational Waves}
\label{sec:matchedfilter}
\begin{figure}[tb]
\centering
\hbox{
\includegraphics[width=\linewidth]{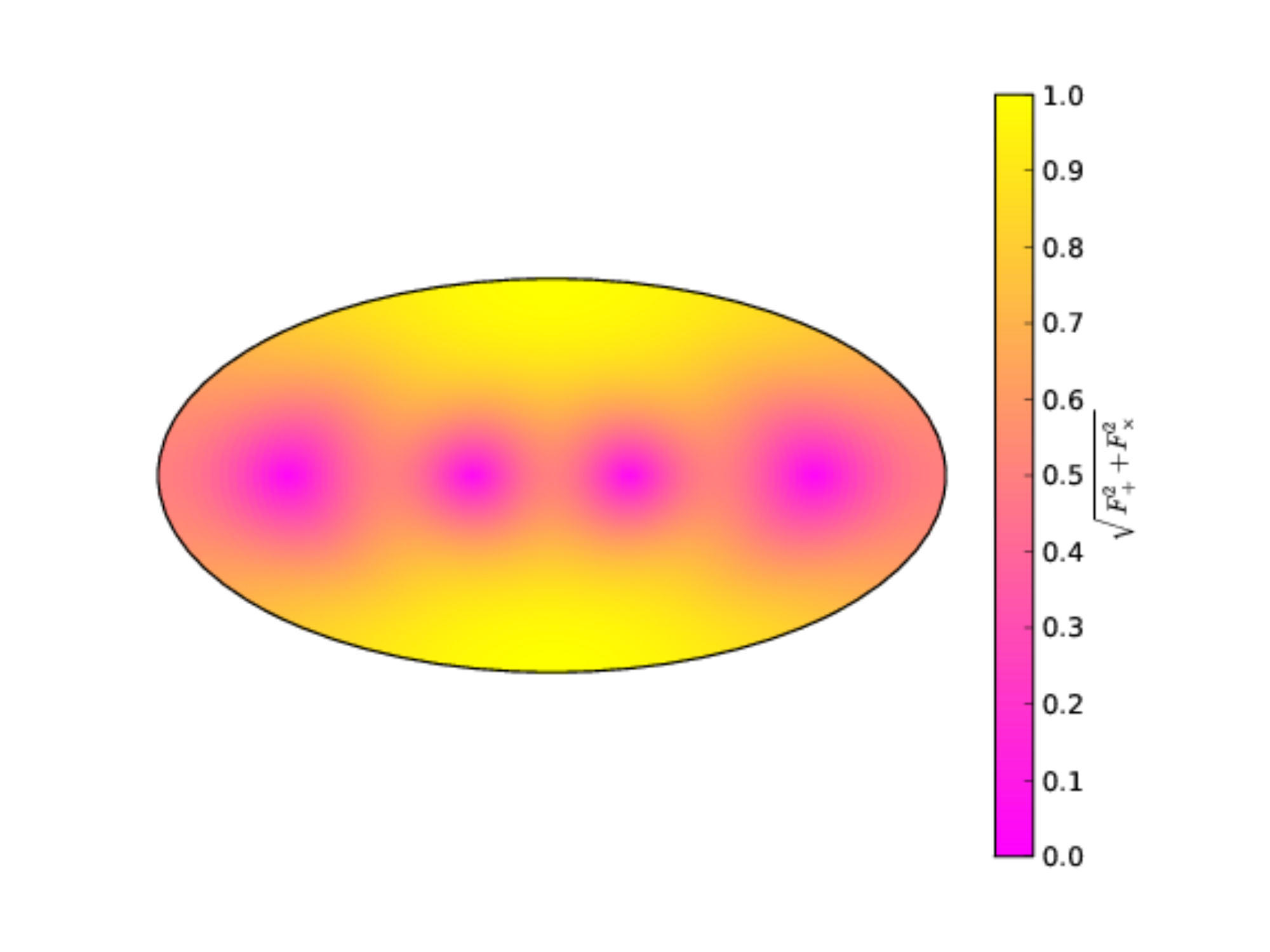}
}
\caption{
Antenna pattern for an interferometric gravitational-wave detector in
source-centric coordinates, $\varphi$ horizontally and $\theta$
vertically.  The arms lie along $\theta=\pi/2,\varphi=0,\pi/2$
respectively.  Such a detector is most sensitive to signals directly
overhead or below, and least sensitive to signals in the plane of the
arms.  The sensitivity drops to zero along the lines between the arms,
$\theta=\pi/2,\varphi=\pm \pi/4$ and $\theta=\pi/2,\varphi=\pm
3\pi/4$.}
\label{fig:antenna}
\end{figure}

The response of an interferometric detector is described by 
an {\it antenna pattern}~\cite{PhysRevD.63.042003},
\begin{align}
\label{eq:AntPattern}
F_+ &= -\frac{1}{2} (1+\cos^2 \theta) \cos 2\varphi \cos 2\psi -
\cos\theta \sin 2 \varphi \sin 2\psi \,, \nonumber \\
F_\times &= \frac{1}{2} (1+\cos^2 \theta) \cos 2\varphi \sin 2\psi -
\cos\theta \sin 2 \varphi \cos 2\psi \,.  \nonumber \\
\end{align}
Following~\cite{Brown:2012gs} we rewrite this in the more convenient form
\begin{align}
\label{eq:AntPatternF0}
F_+ &= F_0 \cos 2(\psi + \psi_0)  \,, \nonumber \\
F_\times &= F_0 \sin 2(\psi + \psi_0)  \,, 
\end{align}
where
\begin{equation}
F_0 = \sqrt{((1+\cos^2\theta)/2)^2 \cos^2 2\varphi + 
\cos^2\theta \sin^2 2\varphi} \nonumber 
\end{equation}
and
\begin{equation}
\tan2\psi_0 = \frac{\cos\theta}{(1+\cos^2\theta)/2} \tan 2\varphi \,. 
\nonumber 
\end{equation}
For reference we show the antenna pattern in fig.(\ref{fig:antenna}).

For gravitational waves, the intrinsic characteristics of a source
are fully encapsulated in the polarization strains $h_+$ and
$h_\times\,.$ When an incoming gravitational wave is incident on the
detector the strains give rise to a signal $s$ given by
\begin{align}
\label{eq:recievedSignal}
&\quad s(\theta,\varphi,\iota,\phi,\psi,t) \nonumber \\
&= F_+(\theta,\varphi,\psi) h_+(\iota,\phi,t)
+ F_\times(\theta,\varphi,\psi) h_\times(\iota,\phi,t) \nonumber \\
&= F_0(\theta, \varphi) h(\psi,\iota,\phi,t) \,,
\end{align}
where we have used eqn.(\ref{eq:AntPatternF0}) and defined
\begin{align*}
h(\psi,\iota,\phi,t) &= \cos 2(\psi + \psi_0) h_+(\iota,\phi,t) \\ 
&+ \sin 2(\psi + \psi_0) h_\times(\iota,\phi,t) \\
\end{align*}

The output of the detector is then $s+n$,
where $n$ is the noise of the detector.  Following standard practice
we incorporate the noise only as $S_n(f)$ and do not add it to the
signal in what follows.  We will take $h$ in
eqn.(\ref{eq:recievedSignal}) to be the output of a numerical
simulation, to be discussed in the following section.

We now briefly review some of the data analysis framework employed in
current LIGO/Virgo searches, and which will be used throughout this
paper.  An inner product on the space of real, time-dependent
waveforms $A(t)$ and $B(t)$, with respect to a given noise curve
described by a power spectral density $S_n(f)$, is
\begin{equation}
\label{eq:InnerProduct}
     \InnerProduct{A(t)|B(t)} 
 = 4\, \mathrm{Re} \int_{0}^\infty df\,
   \frac
     {\tilde{A}(f) \tilde{B}^\star(f)} 
     {S_n(f)} \,.
\end{equation}

In stationary, Gaussian noise, the optimal measure of the presence of
a gravitational wave signal that matches a model waveform, called a
{\it template}, is the \snr{} denoted by $\rho$, with

\begin{equation} 
\label{eq:snr} 
\rho^2 =
\frac{\InnerProduct{s|h_+}^2} {\InnerProduct{h_+|h_+}} \,,
\end{equation}
and where we are studying the response of a single detector to one
polarization, typically taken to be  $h_+$.   We note in passing that
in a multi-detector search the data streams from all instruments will
be filtered against the same $h_+$, and that the source angles
$\iota,\phi$ will be the same at all detectors.  However, the
orientations of the different detectors will provide different values
of $\psi$, making the detectors sensitive to different combinations of
the polarization.  In addition each detector's $F_0$ will have a
different dependence on $\theta, \varphi$ providing coverage of
regions of the sky to which any one detector might be insensitive.

The signal will arrive at an unknown time which we identify as the
time of coalescence and denote $t_0$.  We assume the template waveform
$h$ is a good approximation to the signal $s$, and search for the
signal at all times by shifting the template.  This has the effect in
the Fourier domain of changing $\tilde{h}(f)$ to $\tilde{h}(f)\exp(-2
\pi ift_0)$.  The signal will also have an unknown phase at the time
of coalescence, corresponding to the value of $\phi$ in
fig.(\ref{fig:angles}), which we denote $\phi_0$.  This introduces an
additional factor of $\exp(2 \pi i \phi_0)$.  In practice, this leads
the \snr{} to be evaluated as
\begin{equation}
\label{eq:snrCalc}
\rho(s,h,t_0) =
\frac{4}{\sqrt{\InnerProduct{h_+|h_+}}} \left| \int_0^\infty
\frac{\tilde{s}(f) \tilde{h}_+^\star(f)}{S_n(f)} e^{-2\pi ift_0}\,df
\right|
\end{equation}
where the absolute value removes the dependence on the unknown phase.
Eqn.(\ref{eq:snrCalc}) may be evaluated by a single complex inverse
Fourier transform, and the maximization over $t_0$ is then
accomplished by finding the maximum of the resulting time series.
Eqn.(\ref{eq:snrCalc}) is only an exact calculation of the \snr{} if
$\InnerProduct{h_+|h_\times} = 0$ \cite{Babak:2012zx}, which is not
true in general; however, we expect the errors introduced by this
approximation to be small.

Note that, by eqn.(\ref{eq:recievedSignal}), the dependence on the
\snr{} of the detector angles may be factored out in
eqn.(\ref{eq:snrCalc}).  Note also that $F_0(0,0) = 1$.  These imply
that, given the \snr{} of a signal at $\theta=\varphi=0$, we know the
\snr{} of a signal in the same orientation at all other sky positions.

Related to the \snr{} is the {\it match} or {\it overlap} obtained by
normalizing both waveforms
\begin{equation}
\label{eq:overlap}
\Overlap{s|h_+} 
 = \max_{t_0,\phi_0} \frac{\InnerProduct{s|h_+}}
{\sqrt{ \InnerProduct{s|s} \InnerProduct{h_+|h_+} }}\,.
\end{equation}
The overlap is a measure from 0 to 1 of how well the template matches the signal, an overlap of 1 indicates that the template is an exact match to the signal
and anything lower than one is a diminished match.   

Gravitational-wave strain falls off as the the reciprocal of the
distance between source and detector.  It follows from
eqn.(\ref{eq:snr}) that the \snr{} falls off in the same way, while
the normalization removes the distance dependence of the template.
Henceforth we place the signal $s$ in eqn.(\ref{eq:snr}) at 1 Mpc from
the detector and denote the resulting \snr{} as
$\rho_{1\textrm{Mpc}}$.  We also choose a {\it threshold SNR}, a value
above which indicates the presence of a signal in the data.  We will
take this to be 5.5, the threshold used in current LIGO/Virgo
searches.  The choice of this value is motivated by the behavior of
the noise in the detector~\cite{Allen:2005fk}.  The 
distance at which a signal would have an \snr{} of 5.5 is then
\begin{equation}
\label{eq:snrRatio}
r = \frac{\rho_{1\textrm{Mpc}}}{5.5}
\end{equation}

We now consider two templates, $h_{ideal}$ which exactly matches the
signal and $h$ which in some way approximates the signal.  We can 
determine the fraction of the available distance that is lost by using 
the approximate template as:
\begin{equation}
\frac{r}{r_{ideal}}
=
\frac{\rho_{1\textrm{Mpc}}(h)/5.5}{\rho_{1\textrm{Mpc}}(h_{ideal})/5.5} 
= \frac{\Overlap{s|h}}{\Overlap{s|h_{ideal}}} 
= \Overlap{s|h} \,.
\label{eqn:allRatios}
\end{equation}
The first equality follows from eqn.(\ref{eq:snrRatio}), the second
from dividing both numerator and denominator by the common factor
$\InnerProduct{s|s}^{1/2}$ and the third from the fact that when the
template exactly matches the signal the overlap is 1.  The overlap
therefore measures the fraction of the \snr{} lost by using an
incorrect template, and equivalently the fraction of the distance
lost.  As the universe is approximately uniform at distances accessible
to even initial LIGO~\cite{PhysRevD.85.082002}, the event rate is
approximately equal to the cube of the range, although this will also
depend on the antenna pattern.  However, we note that the overlap does
not give the value of $r_{ideal}$.  As an extreme example, if
$r_{ideal}$ is sufficiently small that the number of expected events
per observation time is close to zero, then the fractional loss of
range implied by a low overlap is inconsequential.

\section{The Binary Black Hole Coalescence Waveforms}
\label{sec:bbhWaveforms}
This paper uses \nr{} waveforms covering the late inspiral, merger and ringdown for a variety of mass ratios and spins. 
All of the  \nr{} simulations used in this study were produced with
GATech's \textsc{Maya} code~\cite{Haas:2012bk,Healy:2011ef,Bode:2011xz,Bode:2011tq,Bode:2009mt,Healy:2009zm}. The
\textsc{Maya} code uses the \texttt{Einstein Toolkit} \cite{et-web}
which is based on the \texttt{CACTUS} \cite{cactus-web} infrastructure
and \texttt{CARPET} \cite{Schnetter-etal-03b} mesh refinement. 
We use sixth-order spatial finite differencing and extract the waveforms 
at a finite radius of $75M$, where $M$ is a code unit set to unity 
and can be scaled to any physical mass scale. All grids have 10 levels of 
refinements unless noted below. 

We use 28 simulations in this paper 
and group them according to their initial parameters in Table \ref{tab:NR}.  
Grid details, including outer boundary and resolution on the finest are
also shown.  The simulations can be separated into three groups: non-spinning,
equal-mass with aligned spin, or unequal-mass with precessing spin.  
For the simulations with $q>4$, we
used the coordinate-dependent gauge term as described in Refs.
\cite{Muller:2010zze} and \cite{Schnetter:2010cz}.  For the $q=10$ and
$q=15$ simulations, initial parameters in Ref. \cite{Nakano:2011pb}
were used. These simulations ($q>4$) have an extra level of refinement 
for 11 levels total, with the exception of $q=6$ and $q=15$.  
These have 10 levels and 
12 levels, respectfully.

The output of all simulations is the Weyl Scalar, $\Psi_4$,
decomposed into spin-weighted spherical harmonics.  Simulations are
performed in a coordinate system which we will denote the {\it
source-centric} frame, to distinguish it from the {\it
detector-centric} frame we will employ subsequently.  See
fig.(\ref{fig:angles}) for the definition of the angles used in this
frame.  In terms of these angles the decomposition is:
\begin{equation}
\label{eq:Psi4Decomposition}
r M \Psi_4(\iota,\phi,t) = \sum_{l,m} {}_{-2} Y_{\ell m}(\iota,\phi) C_{\ell m}(t)\,.
\end{equation}
This is related to the strain measured by gravitational-wave
observatories as
\begin{eqnarray}
\label{eq:StrainDecomposition}
 \Psi_4(\iota,\phi,t) &=& -(\ddot{h}_+(\iota,\phi,t) - i
\ddot{h}_\times(\iota,\phi,t)) \nonumber \\
&=& \sum_{\ell m} {}_{-2} Y_{\ell m}(\iota,\phi) \ddot{h}^\star_{\ell m}(t)\,.
\end{eqnarray}
The quadrupole mode is given by $(\ell,|m|)=(2,2)\,.$
Throughout this paper we work in the frequency domain, and 
therefore avoid the integration  to strain since 
$\tilde{h} = \tilde{\Psi}_4 / (-4\pi^2f^2)$.

\begin{table*}
  \begin{tabular}{|r|c|c|c|c|c|c|c|c|c|}
\hline
ID & $q$  & $m_{b+}/M$ & $m_{b-}/M$ & $\chi_+$ & $\chi_+$ & $p_+/M$ & $d/M$ & $R_{b}/M$ & $M/h_{fine}$\\
\hline
Q01 & 1.00 & 0.485923 & 0.485923 & 0.0 & 0.0  & (-0.00098038, 0.096107, 0) & 10.00 & 317.44 & 103 \\
Q02 & 1.15 & 0.520973 & 0.451009 & 0.0 & 0.0  & (-0.00097306, 0.095648, 0) & 10.00 & 317.44 & 103 \\
Q03 & 1.30 & 0.551561 & 0.420763 & 0.0 & 0.0  & (-0.00095146, 0.094500, 0) & 10.00 & 317.44 & 103 \\
Q04 & 1.45 & 0.578486 & 0.394310 & 0.0 & 0.0  & (-0.00092318, 0.092922, 0) & 10.00 & 317.44 & 103 \\
Q05 & 1.50 & 0.586758 & 0.386214 & 0.0 & 0.0  & (-0.00091215, 0.092328, 0) & 10.00 & 317.44 & 103 \\
Q06 & 1.60 & 0.602367 & 0.370978 & 0.0 & 0.0  & (-0.00088915, 0.091074, 0) & 10.00 & 317.44 & 103 \\
Q07 & 1.75 & 0.623691 & 0.350248 & 0.0 & 0.0  & (-0.00085215, 0.089076, 0) & 10.00 & 317.44 & 103 \\
Q08 & 1.90 & 0.642849 & 0.331709 & 0.0 & 0.0  & (-0.00081702, 0.086999, 0) & 10.00 & 317.44 & 103 \\
Q09 & 2.00 & 0.654574 & 0.320400 & 0.0 & 0.0  & (-0.00079295, 0.085598, 0) & 10.00 & 317.44 & 103 \\
Q10 & 2.05 & 0.660153 & 0.315030 & 0.0 & 0.0  & (-0.00078063, 0.084896, 0) & 10.00 & 317.44 & 103 \\
Q11 & 2.20 & 0.675859 & 0.299945 & 0.0 & 0.0  & (-0.00074412, 0.082799, 0) & 10.00 & 317.44 & 103 \\
Q12 & 2.35 & 0.690180 & 0.286237 & 0.0 & 0.0  & (-0.00070983, 0.080733, 0) & 10.00 & 317.44 & 103 \\
Q13 & 2.50 & 0.703291 & 0.273726 & 0.0 & 0.0  & (-0.00067707, 0.078713, 0) & 10.00 & 317.44 & 103 \\
H01 & 1.00 & 0.487207 & 0.487207 & 0.0 & 0.0  & (-0.00071204, 0.090099, 0) & 11.00 & 409.60 & 200 \\
H02 & 2.00 & 0.655683 & 0.321576 & 0.0 & 0.0  & (-0.00057168, 0.080204, 0) & 11.00 & 409.60 & 200 \\
H03 & 3.00 & 0.740897 & 0.239917 & 0.0 & 0.0  & (-0.00041607, 0.067799, 0) & 11.00 & 409.60 & 200 \\
H04 & 4.00 & 0.792317 & 0.191313 & 0.0 & 0.0  & (-0.00030795, 0.057941, 0) & 11.00 & 409.60 & 200 \\
H05 & 5.00 & 0.826040 & 0.158317 & 0.0 & 0.0  & (-0.00033261, 0.053831, 0) & 10.00 & 409.60 & 240 \\
H06 & 6.00 & 0.850747 & 0.135461 & 0.0 & 0.0  & (-0.00026264, 0.047519, 0) & 10.00 & 409.60 & 200 \\
H07 & 7.00 & 0.869309 & 0.118371 & 0.0 & 0.0  & (-0.00021252, 0.042488, 0) & 10.00 & 409.60 & 320 \\
H08 & 10.00& 0.907397 & 0.085237 & 0.0 & 0.0  & (-0.00016852, 0.036699, 0) &  8.39 & 409.60 & 400 \\
H09 & 15.00& 0.936224 & 0.057566 & 0.0 & 0.0  & (-0.00016052, 0.029072, 0) &  7.25 & 409.60 & 800 \\
\hline
S01 & 1.00 & 0.453711 & 0.453711 & -0.4 & -0.4& (-0.00079326, 0.092237, 0) & 11.00 & 409.60 & 200 \\
S02 & 1.00 & 0.453865 & 0.453865 & 0.4 & 0.4  & (-0.00065074, 0.088023, 0) & 11.00 & 409.60 & 200 \\
S03 & 1.00 & 0.303458 & 0.303458 & 0.8 & 0.8  & (-0.00060332, 0.086010, 0) & 11.00 & 409.60 & 200 \\
\hline
P01 & 4.00 & 0.655334 & 0.156900 & (0.6, 0.0, 0.0)   & (-0.6, 0.0, 0.0) & (0, 0.066502, 0) & 9.00 & 409.60 & 140 \\
P02 & 4.00 & 0.655306 & 0.156762 & (0.3, 0.0, -0.5)  & (-0.6, 0.0, 0.0) & (0, 0.068787, 0) & 9.00 & 409.60 & 140 \\
P03 & 4.00 & 0.655306 & 0.156764 & (-0.3, 0.0, -0.5) & (-0.6, 0.0, 0.0) & (0, 0.068758, 0) & 9.00 & 409.60 & 140 \\
\hline
\end{tabular}

\caption{\label{tab:NR}\textbf{Simulations Used}: The 28 simulations' initial parameters and grid structures are listed. The table is split into three groups: non-spinning, equal-mass with spin, and precessing spins. The table contains $q=m_+/m_-$, the bare puncture masses $m_{b+}/M$ and $m_{b-}/M$, the non-dimensional spins, $\chi_i = S_i/m_i^2$, the initial momentum, $p_+/M$, the initial separation, $d/M$, the outer boundary, $R_b/M$, and the resolution on the finest refinement level $M/h_{fine}$.  If only one spin value is listed, the spin is aligned with the initial angular momentum. }
\end{table*}

\section{Overlap}
\label{sec:preliminaryInvestigations}
We start by examining the relative importance of the non-dominant 
modes in a waveform comparison.  The full waveform involves factors of the spherical
harmonics and the amplitudes of the modes (see eqn.(\ref{eq:Psi4Decomposition})).
When the amplitudes of the higher modes are vanishingly
small, they can be ignored; however, as we have already noted in fig.(\ref{fig:modeAmplitudes}),
the relative amplitudes grow in strength with mass ratio.

\begin{figure}[tb]
\centering
\hbox{
\includegraphics[width=.45\linewidth]{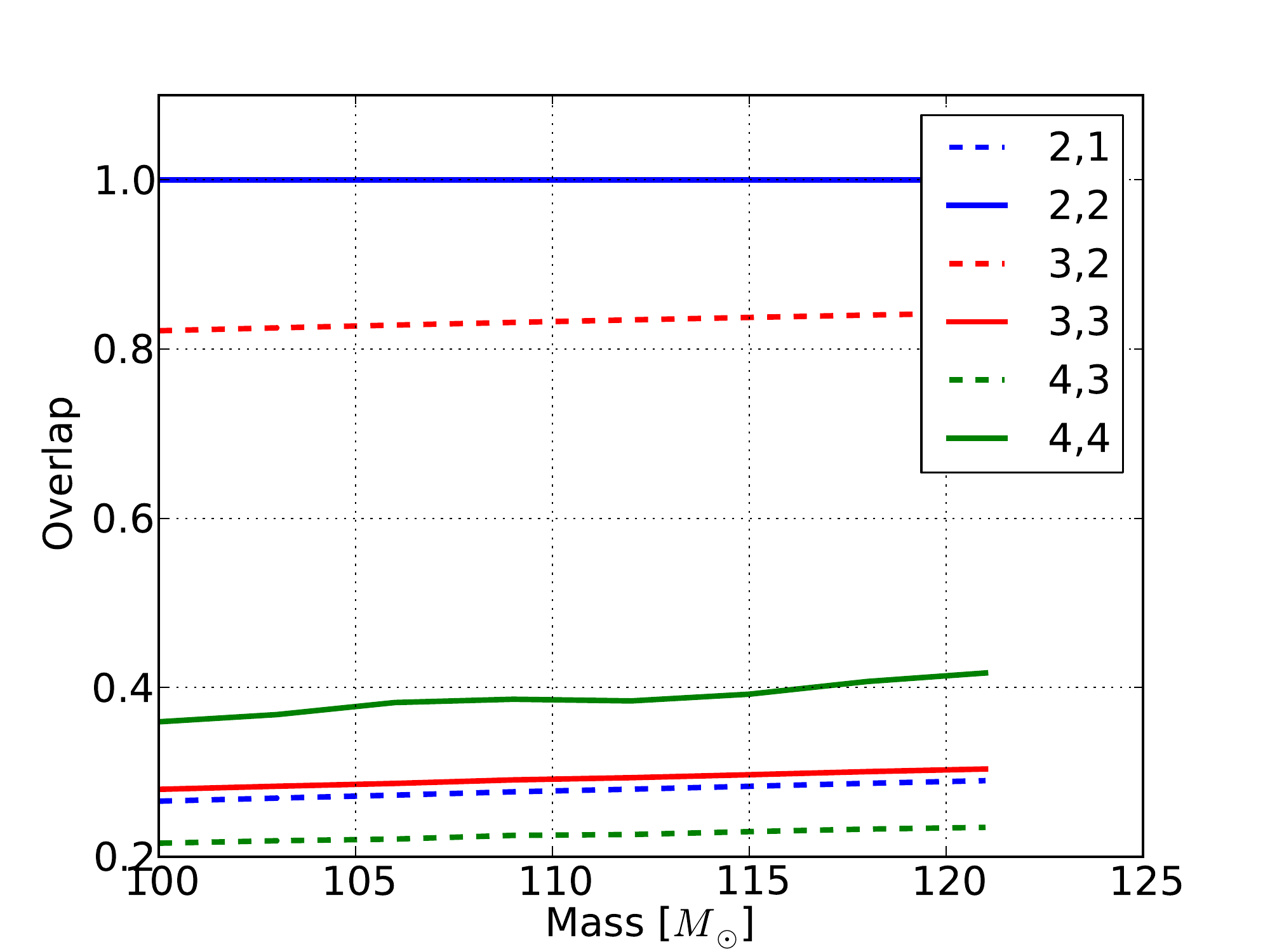}
\includegraphics[width=.45\linewidth]{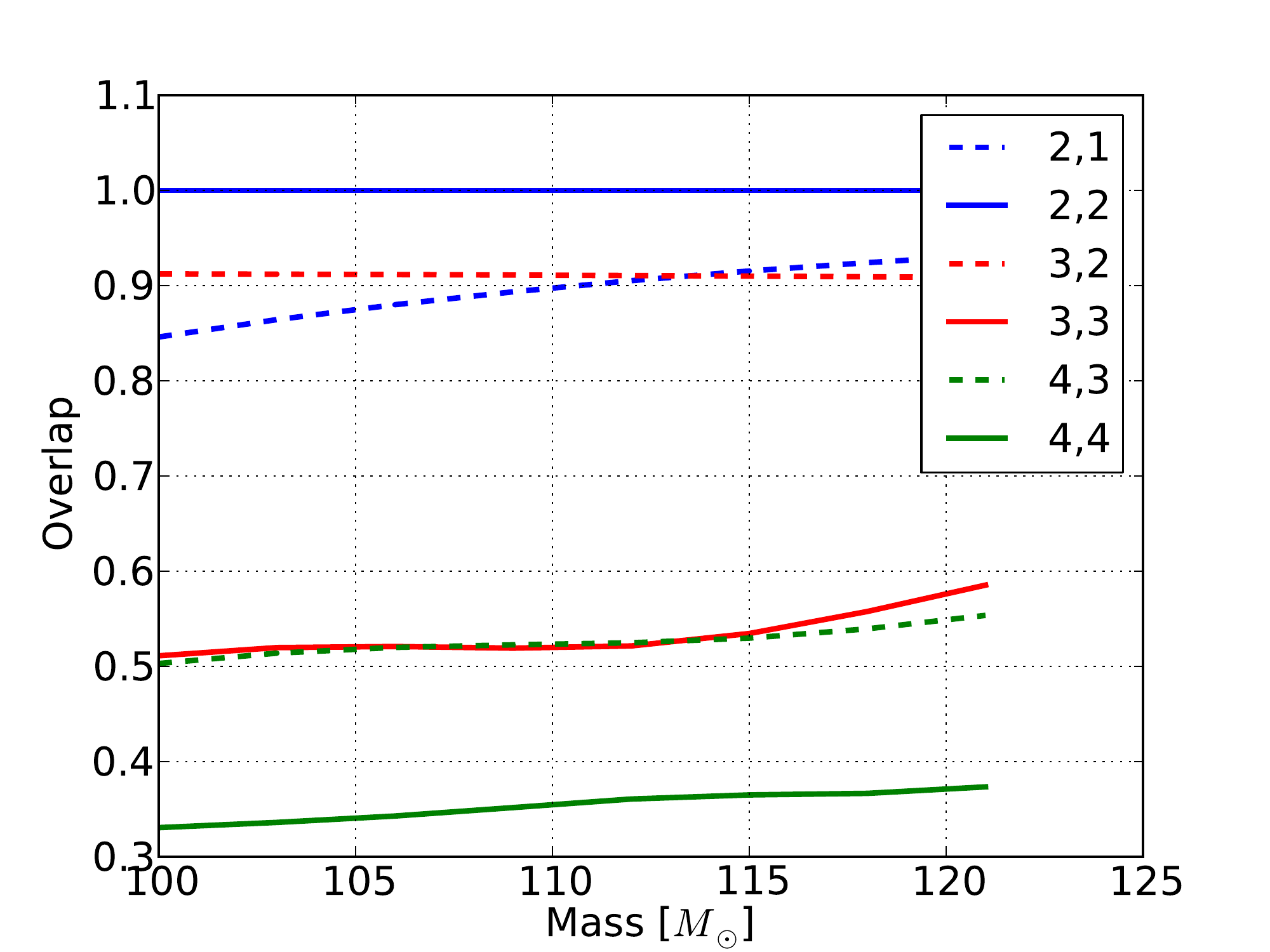}
}
\caption{Overlap of higher modes with 2,2 for 
{\bf Left}: $q=1$ and {\bf Right}: $q=4$ systems.  In both cases the
most significant higher modes have poor overlaps with (2,2),
suggesting that a $h_{22}$ will be a poor fit to the full signal in
regions dominated by these modes.
}
\label{fig:modeOverlaps}
\end{figure}

\begin{figure}[h]
\centering
\hbox{
\includegraphics[width=\linewidth]{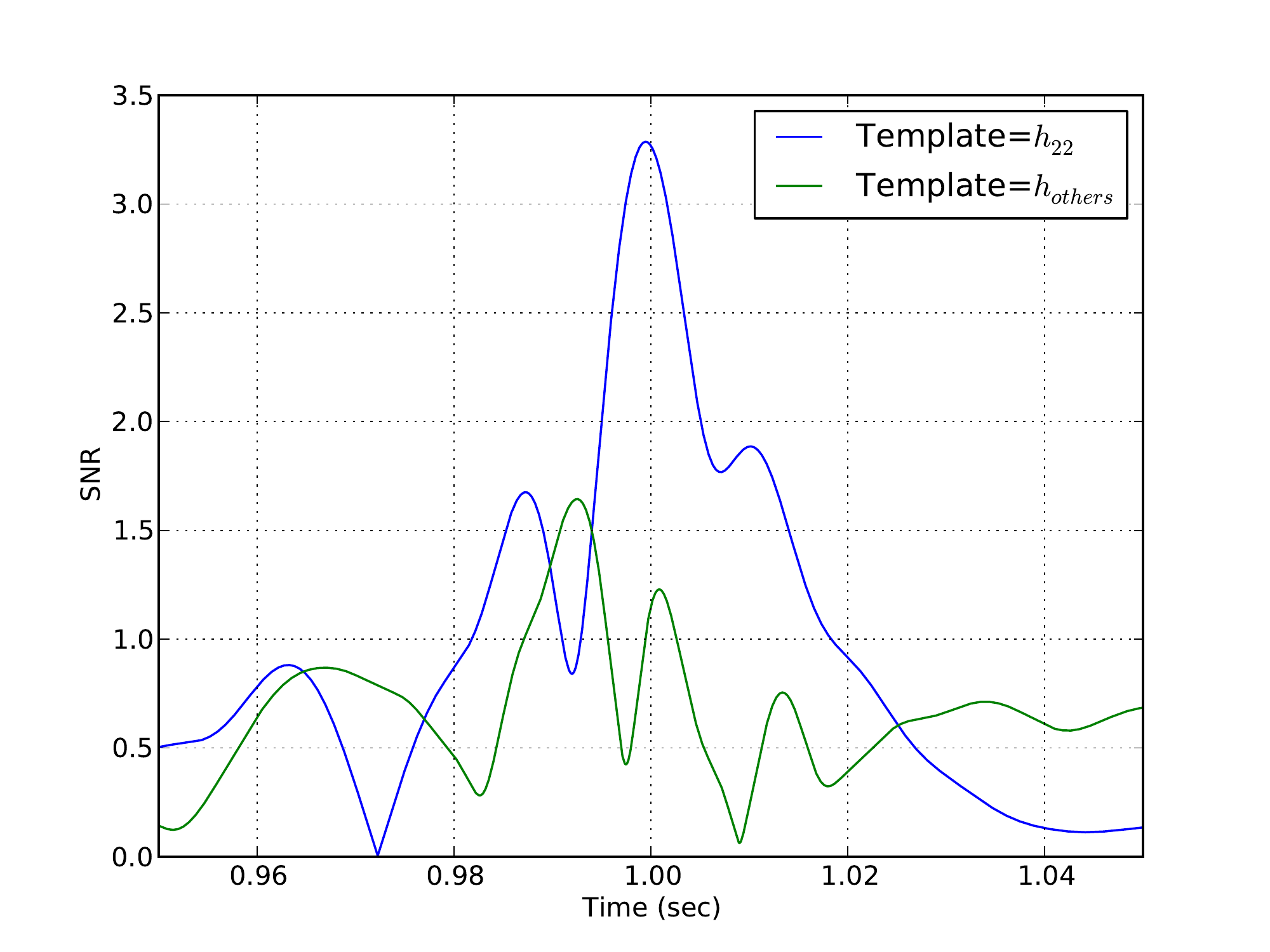}
}
\caption{
\snr{} time series for $\rho(s,h_{2,2})$ and $\rho(s,h_{others})$.  The
specific behavior will depend on the angles, the values here were
chosen to illustrate the issue, $\theta=2.36, \varphi=2.58, \iota=1.54,
\phi=5.16$.  At the time when the $h_{2,2}$ series peaks, $h_{others}$
has dropped by 38\%.  The tension between the modes means that the
total \snr{} will be less than the sum of the component \snr{}s.
}
\label{fig:snrTimeSeries}
\end{figure}

In fig.(\ref{fig:modeOverlaps}) we plot the overlap of each mode
against (2,2) individually.  If all modes matched well against (2,2)
it would suggest that a template containing only this mode would
be a good match to the full signal, regardless of the source
orientation; however, we find that not to be the case.  In both the
$q=1$ and $q=4$ case, the overlap between (2,2) and the next
most dominant modes is poor, below 0.6.  Furthermore, although the
inner product, eqn.(\ref{eq:InnerProduct}), and the decomposition into
modes, eqn.(\ref{eq:Psi4Decomposition}), are themselves linear, the maximization
over time and phase introduces nonlinearities.  In particular,
defining $h_{others} = \sum_{l,m \neq 2,2} h_{lm}$, the sum is only linear if the inner products maximize at the same time.
If not, there will
be a ``tension'' in the modes and the combined \snr{} will be less than
the sum of the individual \snr{}s, i.e. 
\begin{equation}
\rho^2(s, h) \neq \rho^2(s,h_{22}) + \rho^2(s,h_{others}) \,.
\label{eqn:snrSimplify}
\end{equation}
 To quantify this we plot the time
series of both \snr{}s on the right-hand side of eqn.(12) in
fig.(\ref{fig:snrTimeSeries}).  The two series peak at notably
different times, and at the peak of the $h_{22}$ series the
$h_{other}$ series has dropped by 38\% thus we can conclude that the
non-linearities are important, and we cannot use the linear approximation.

\begin{figure}[tb]
\centering
\hbox{
\includegraphics[width=.5\linewidth]{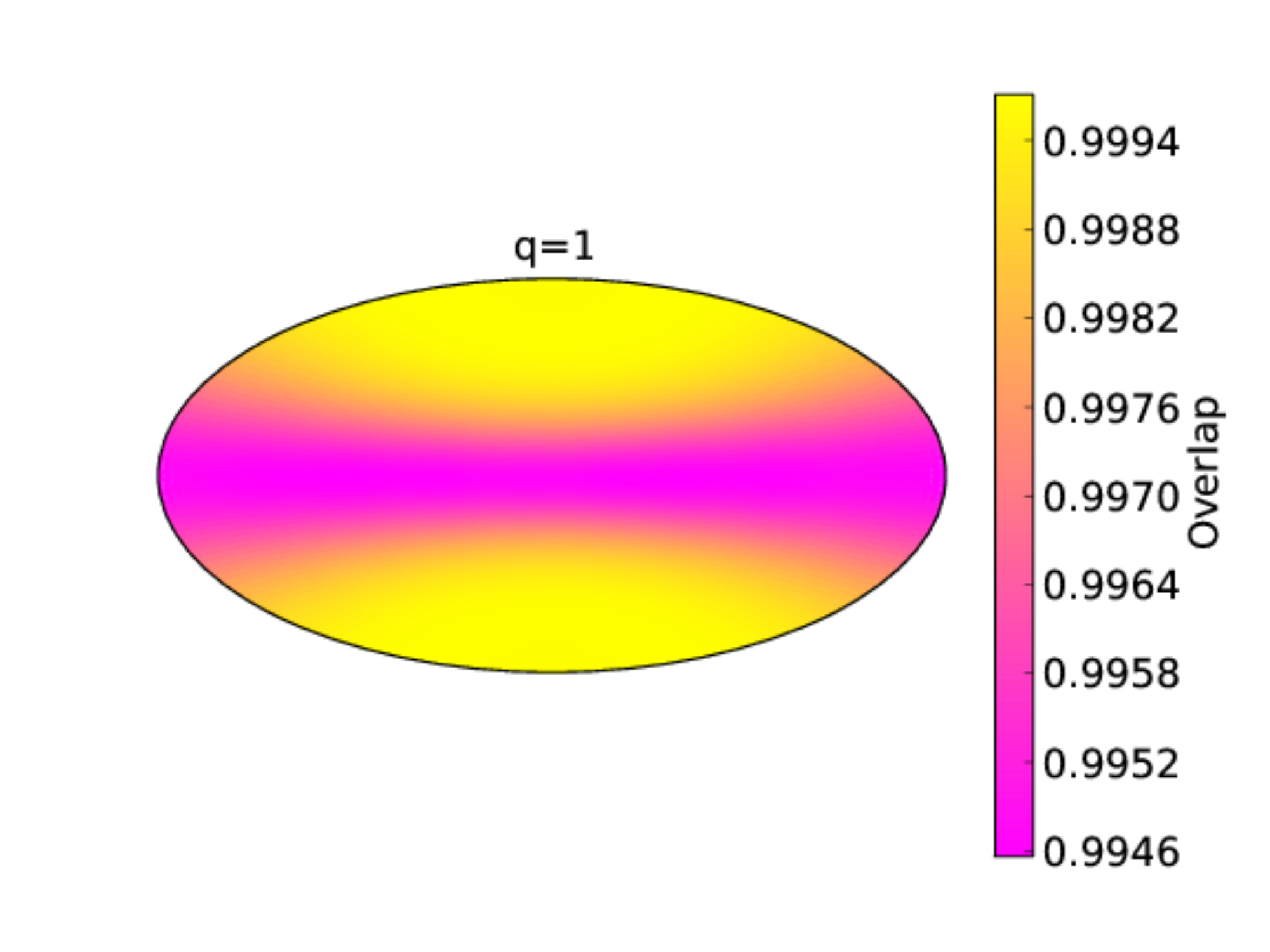}
\includegraphics[width=.5\linewidth]{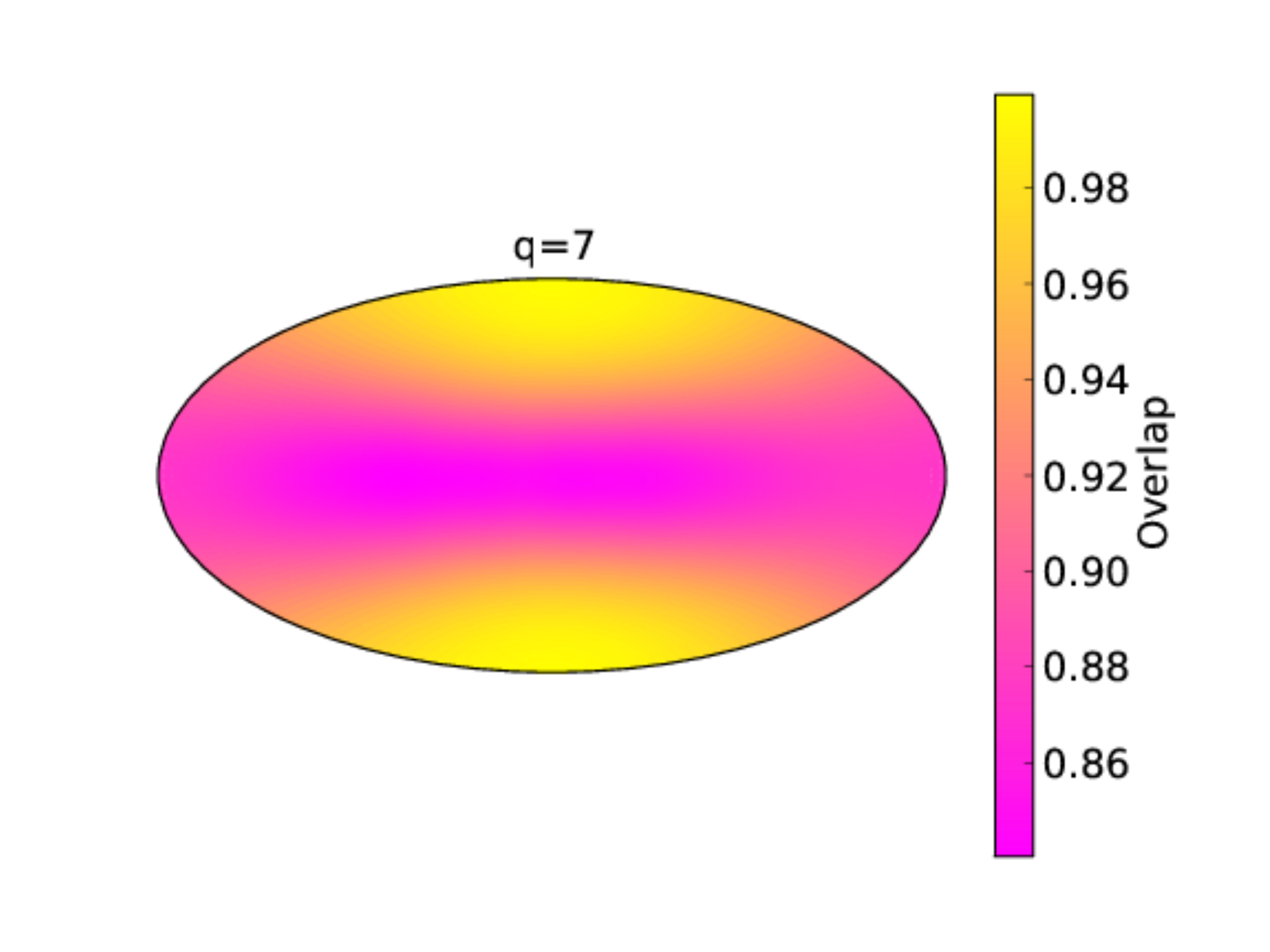}
}
\hbox{
\includegraphics[width=.5\linewidth]{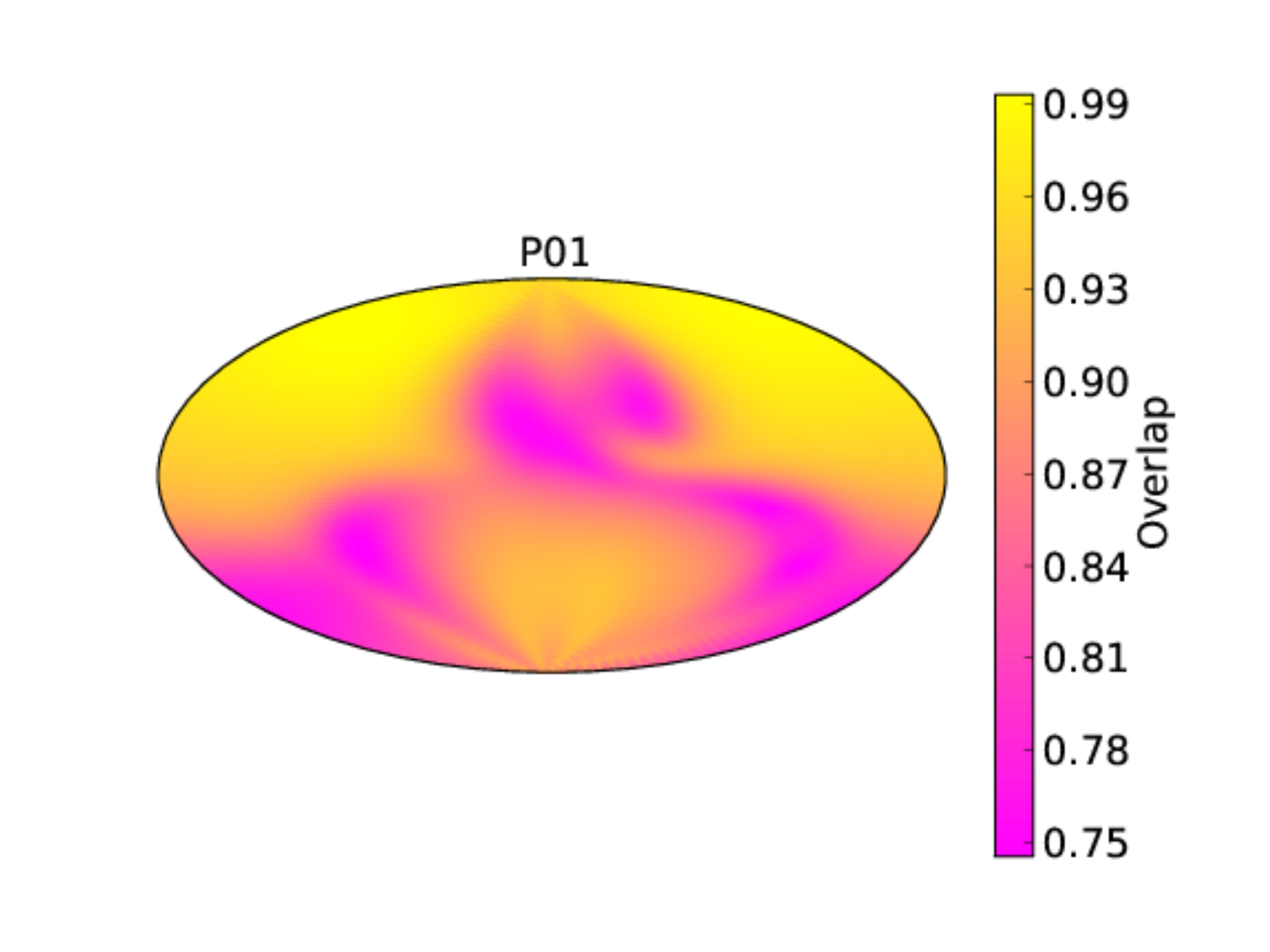}
\includegraphics[width=.5\linewidth]{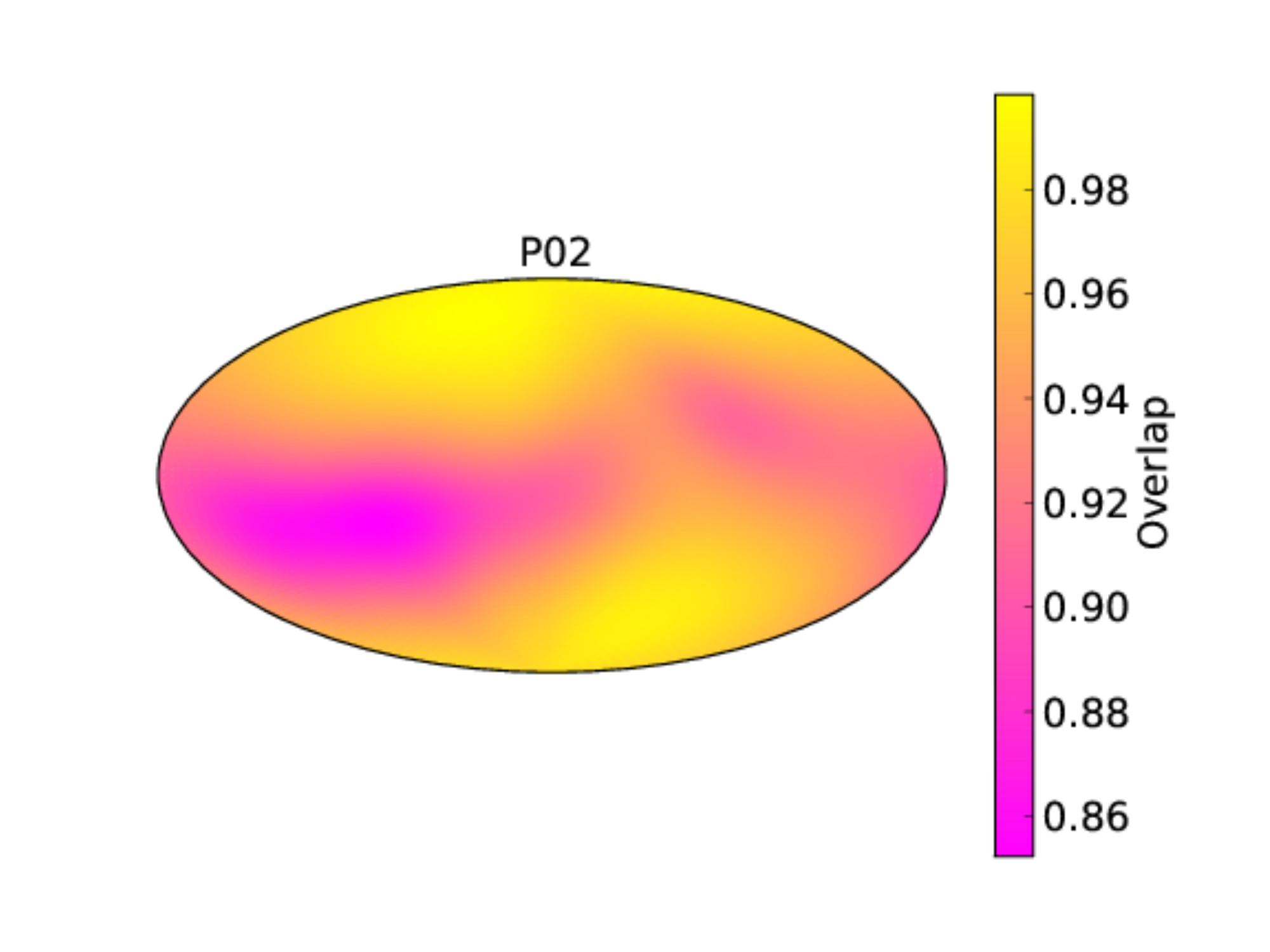}
}
\hbox{
\includegraphics[width=.5\linewidth]{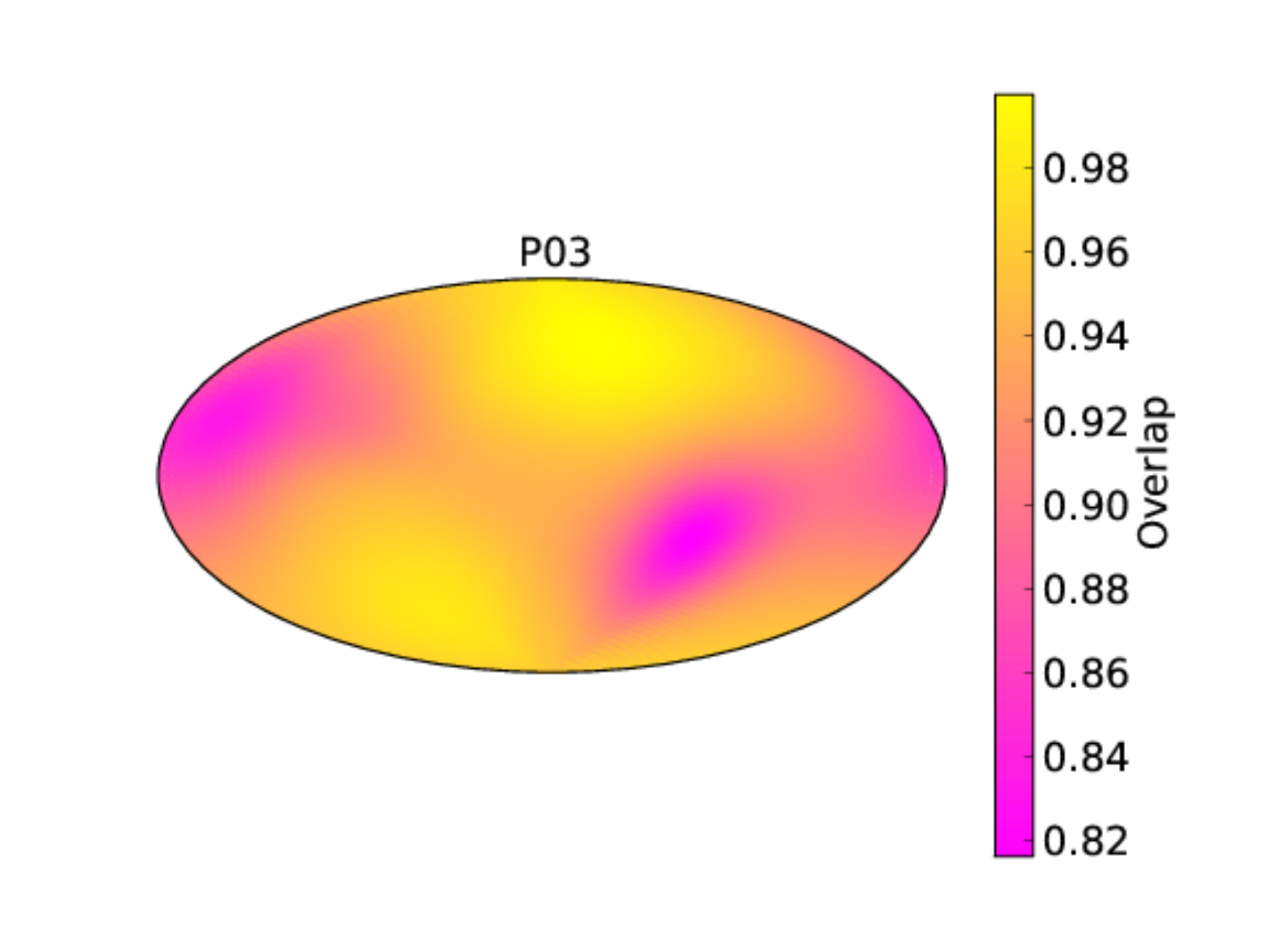}
}
\caption{Overlaps 
in source-centric coordinates, $\phi$ horizontally and 
$\iota$ vertically,
between the complete waveform and the $(2,2)$ mode
for {\bf Top:} the non-spinning $q=1$ and $q=4$,
{\bf Middle} the precessing P01 and P02 and 
{\bf Bottom:} the precessing P03 signals
from tab.(\ref{tab:NR}).  The general features of the non-spinning
images are representative of all mass ratios and (anti-) aligned spin
systems; overlaps are 1.0 at $\iota=0, \pi$ where the full signal
reduces to the (2,2) mode, and are lowest at $\iota=\pi/2$.  There is
more interesting structure in the precessing cases. 
}
\label{fig:sourceCentricOverlaps}
\end{figure}

While fig.(\ref{fig:modeOverlaps}) shows that the (2,2) mode is not an effective 
representation of the other modes,  how well does the (2,2) mode cover the sky of the source?
The overlap between the full-mode waveform and the (2,2)
mode is a function of the angles centered at the source, $(\iota,\phi)$.  The
(2,2)-only template depends on the angles through a single factor,
${}_2Y_{22}(\iota,\phi)$, which is canceled by the normalization; and, therefore, 
we simplify the overlap by placing this waveform at $\iota=\phi=0$.  We
also place both waveforms optimally in the sky of the detector, at
$\theta=\varphi=0$, and choose $\psi=0$.  We will generalize this
momentarily.  Fig.(\ref{fig:sourceCentricOverlaps}) shows the
resulting overlaps for five cases, the non-spinning $q=1$ and $q=7$,
and the precessing cases from tab.(\ref{tab:NR}).  At $\iota=0,\pi$
the waveform is dominated by the $(2, 2)$ modes, the overlap 
approaches 1.0 at these points.  Equation~(\ref{eqn:allRatios}) then
implies that there is no loss of distance incurred by searching with
the (2,2)-only template for systems that are oriented face-on with
respect to the detector.  We can further quantify this by determining
the faction of surface area over which the overlap falls below 0.97\%,
where this value is motivated by the allowed 3\% loss of \snr{} from
using a discrete set of templates~\cite{PhysRevD.85.082002}.
Tab.(\ref{tab:overlapSummaries}) lists this value for several
simulations, along with the the average, median and lowest overlaps as
further measures of the impact of the higher modes.

\begin{table}
  \begin{center}
    \begin{tabular}{|r|r|r|r|r|r|r|}
      \hline
      ID & $q$ & $a$ & \% of area  & Average & Median & Minimum \\
         &     &     & $\geq$ 0.97 &         &        & \\
      \hline
      H01 &  1 &  0          & 100 & 0.997 & 0.998 & 0.995 \\
      H03 &  3 &  0          &  51 & 0.955 & 0.951 & 0.918 \\
      H04 &  4 &  0          &  43 & 0.937 & 0.931 & 0.885 \\
      H05 &  5 &  0          &  40 & 0.927 & 0.920 & 0.868 \\
      H06 &  6 &  0          &  37 & 0.916 & 0.907 & 0.847 \\
      H07 &  7 &  0          &  36 & 0.907 & 0.898 & 0.840 \\
      H08 & 10 &  0          &  36 & 0.903 & 0.892 & 0.826 \\
      H09 & 15 &  0          &  35 & 0.897 & 0.886 & 0.817 \\
      \hline
      S01 &  1 & -0.4        & 100 & 0.997 & 0.997 & 0.993 \\
      S02 &  1 &  0.4        & 100 & 0.997 & 0.997 & 0.994 \\
      S03 &  1 &  0.8        & 100 & 0.997 & 0.997 & 0.994 \\
      \hline
      P01 &  4 & 0.6 (90${}^\circ$)  &  13 & 0.883 & 0.889 & 0.741 \\
      P02 &  4 & 0.6 (150${}^\circ$) &  41 & 0.938 & 0.939 & 0.852 \\
      P03 &  4 & 0.6 (210${}^\circ$) &  28 & 0.933 & 0.942 & 0.816 \\
     \hline
   \end{tabular}
  \end{center}
  \caption[TOADD]{
   \label{tab:overlapSummaries} Summary values of the overlaps between
the (2,2) mode and the full template as a function of the orientation
angles $(\iota,\phi)$.  Names in parenthesis refer to
tab.(\ref{tab:NR}).  Note that the P01 precessing system has lower
overlaps, and a smaller fraction of overlaps greater than 0.97, then
the other systems.
}
\end{table}

Figures (5-7) and tab.~(\ref{tab:overlapSummaries}), all tell the same
story for a single detector when the intrinsic parameters are kept fixed to the signal: the q=1 case is well served with a (2,2)-only waveform over all
source angles.  The higher the mass ratio, the worse a (2,2)-only
waveform does in matching the signal, and this fraction of angles over
which the match does poorly increases.  Furthermore, a precessing
system is badly served by a (2,2) waveform.  We will explore this matter further in future work.

\begin{figure}[tb]
\centering
\hbox{
\includegraphics[width=\linewidth]{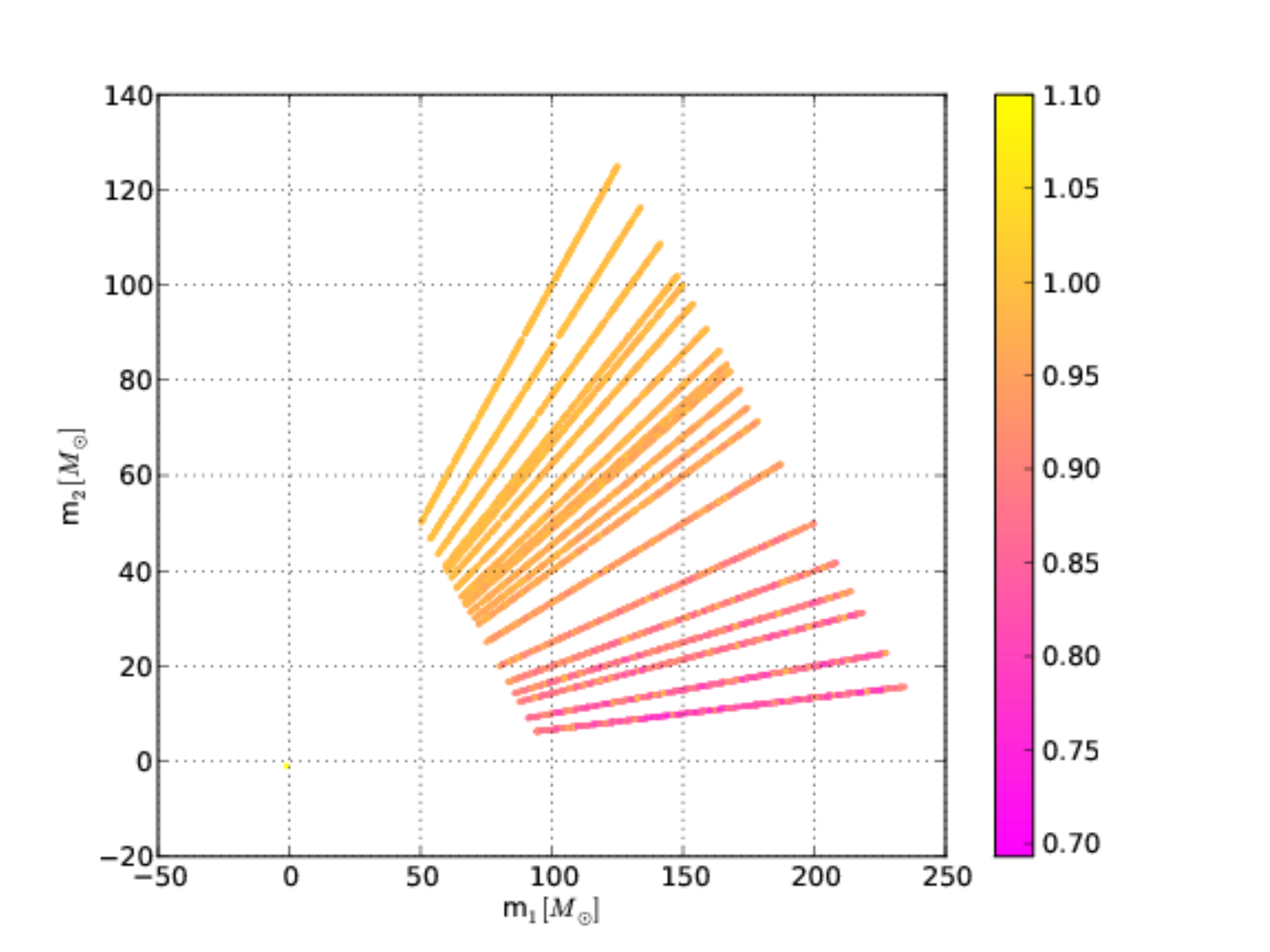}
}
\caption{Overlaps between the complete waveform and 
the $(2,2)$ mode for non-spinning 
waveforms with mass ratios from $1$ to $15$, with all angles and
total mass chosen randomly. At higher mass ratio more of the
total power is distributed into the higher modes and the match
drops accordingly.
}
\label{fig:allOverlaps}
\end{figure}

We now generalize the previous results to
include other values of the detector-centric angles $(\theta,\varphi)$
and $\psi$.   Consider two templates: $h_{22}$ which, as in current
searches, contains only the $(2,2)$ mode of the NR waveform optimally
oriented ($\theta=\varphi=\iota=\phi=\psi=0$), and a perfect template
$h_{ideal}$ which exactly matches the signal.  In
fig.(\ref{fig:allOverlaps}) we show the overlap between the signal and
$h_{22}$ for several non-spinning systems.  Each colored line on the
graph represents a system mass ratio, moving along the line gives
different system masses.  As we move from top to bottom, we are moving
from q=1 to q=15.   The difference in colors along the line give the
overlap value.  The plot shows that for higher mass ratios the
total power is distributed into the higher modes and the match drops
accordingly.  This is consistent with \cite{McWilliams:2010eq,Brown2012}. 

\begin{figure}[tb]
\centering
\hbox{
\includegraphics[width=.45\linewidth]{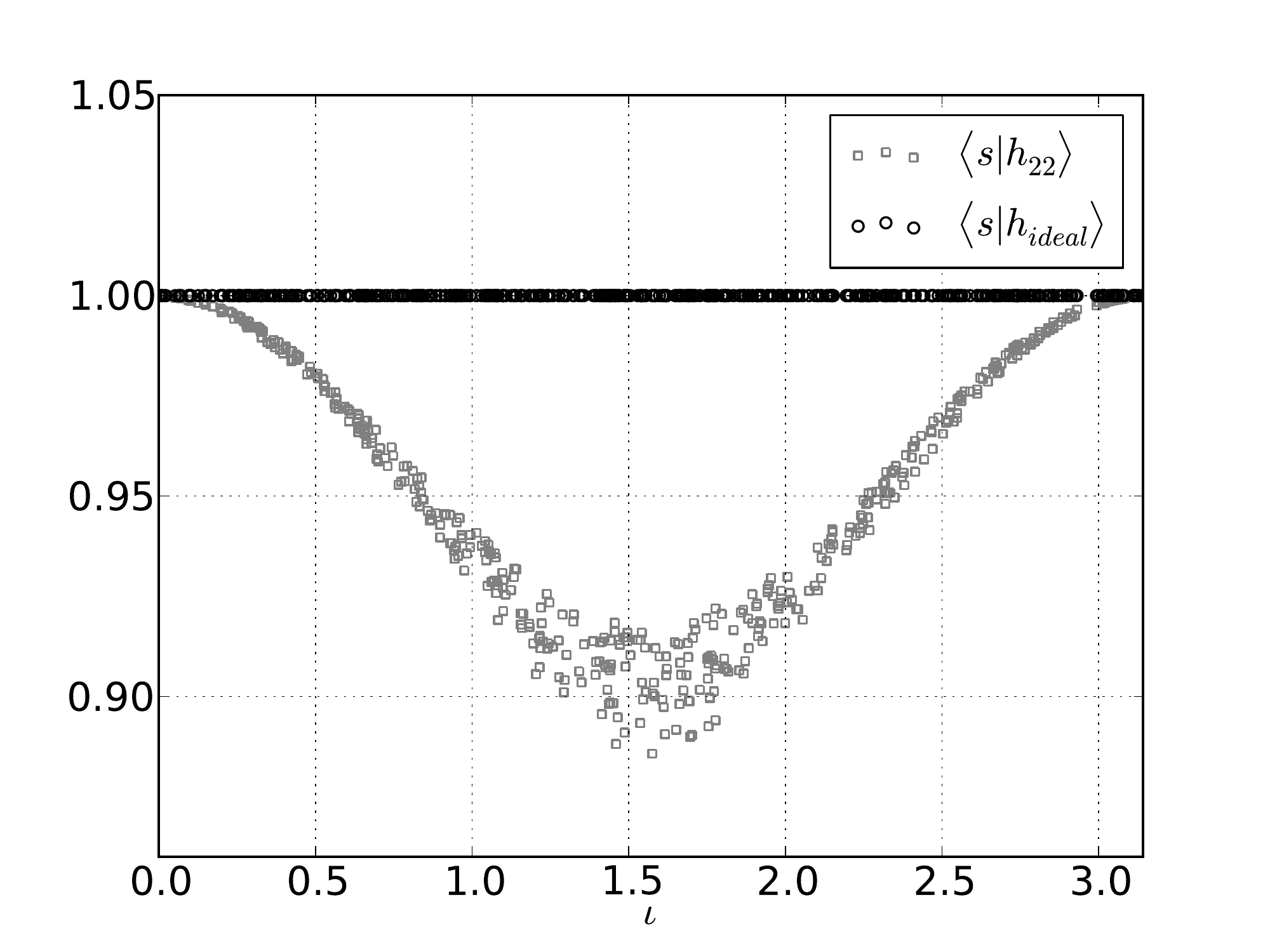}
\includegraphics[width=.45\linewidth]{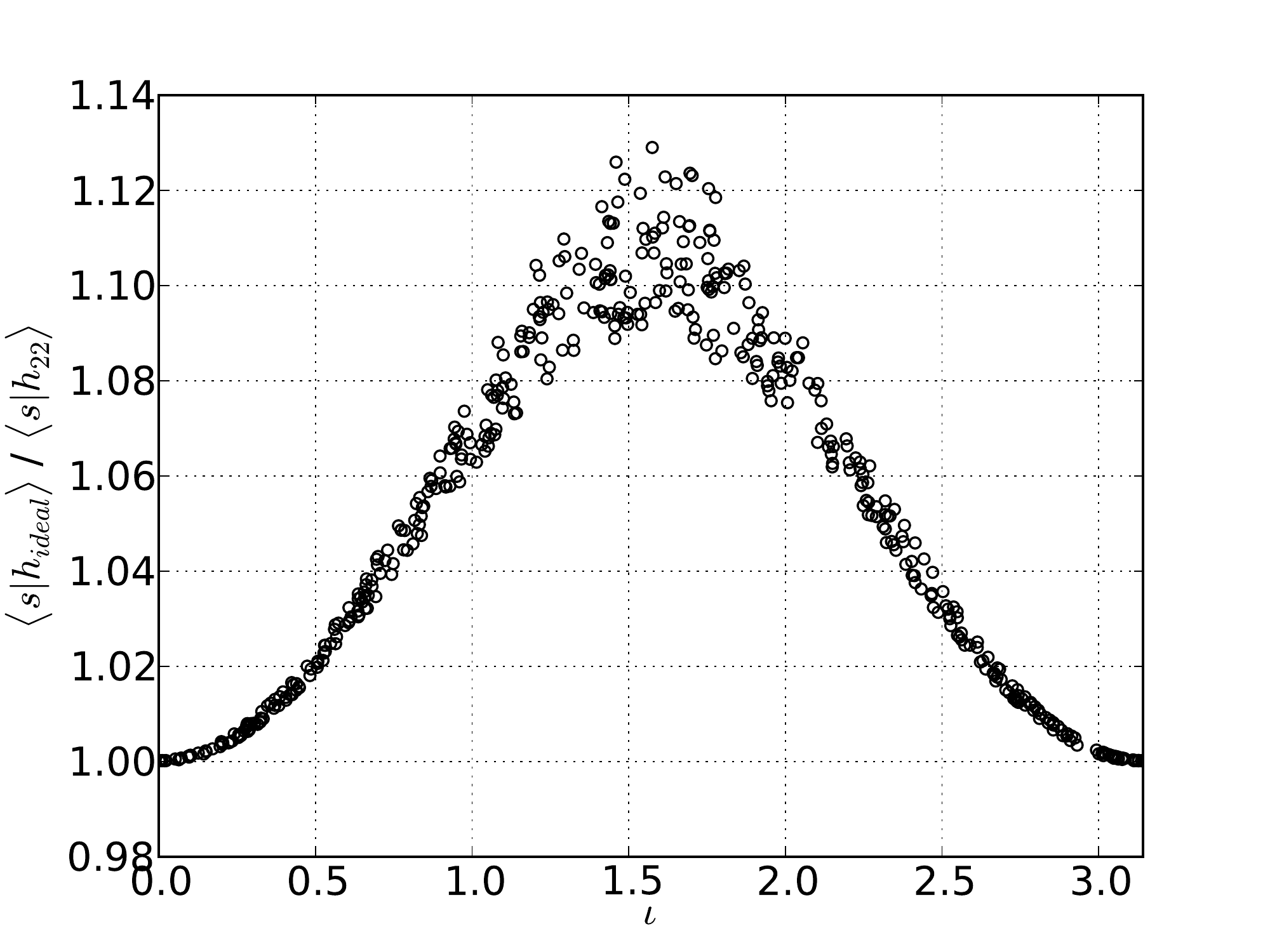}
}
\caption{{\bf Left}: The overlaps obtained using both templates.
Since $h_{ideal}=s$ the overlap is 1.  {\bf Right}: The ratio of the
overlaps.  This is identical to the ratio of SNRs as the additional
factors of $\InnerProduct{s|s}$ cancel.}
\label{fig:overlap_ratio}
\end{figure}

Now consider the $q=4$ non-spinning system, scaled to to $100 M_\odot$
and placed at a distance of 1Gpc from the detector, and examine the
overlap between the signal and both templates.  We randomly choose
values for all angles and plot results with respect to $\iota$, which
has the most significant dependence.  The results are shown in
fig.(\ref{fig:overlap_ratio}), which illustrates that at $\iota=0,\pi$
the variation of the additional angles do not affect the overlap,
while the spread in results widens towards $\iota=\pi/2$.  This again
shows that the (2,2) mode only captures a face-on source orientation
and misses the source as its inclination increases toward the edge-on
case.    This would imply that the higher modes are essential for detecting non-optimally oriented signals, 
but how far away can a single detector see these cases?   We quantify how important
the modes will be in terms of SNR and volume reach in the next section.

\section{SNR and Volume}
\label{sec:results}
As noted at the end of $\S~\ref{sec:matchedfilter}$, the overlap is
equal to the fractional loss in distance to which a signal can be
detected, but this value should be viewed in light of the maximum
possible distance.  This maximum distance depends on three factors:
(1) the total energy radiated by the source, (2) the ability of the
template to extract energy of the signal from the background noise and
(3) the location of the source in the sky of the detector.  For
example, in the plane of the detector along the lines 45 degrees to
the arms, the response goes to zero.  Along these lines the loss in
range implied by a low overlap is irrelevant for a single detector.
In this section we consider the accessible distances, noting the
influence of all three factors.  

\begin{figure}[tb]
\centering
\hbox{
\includegraphics[width=.5\linewidth]{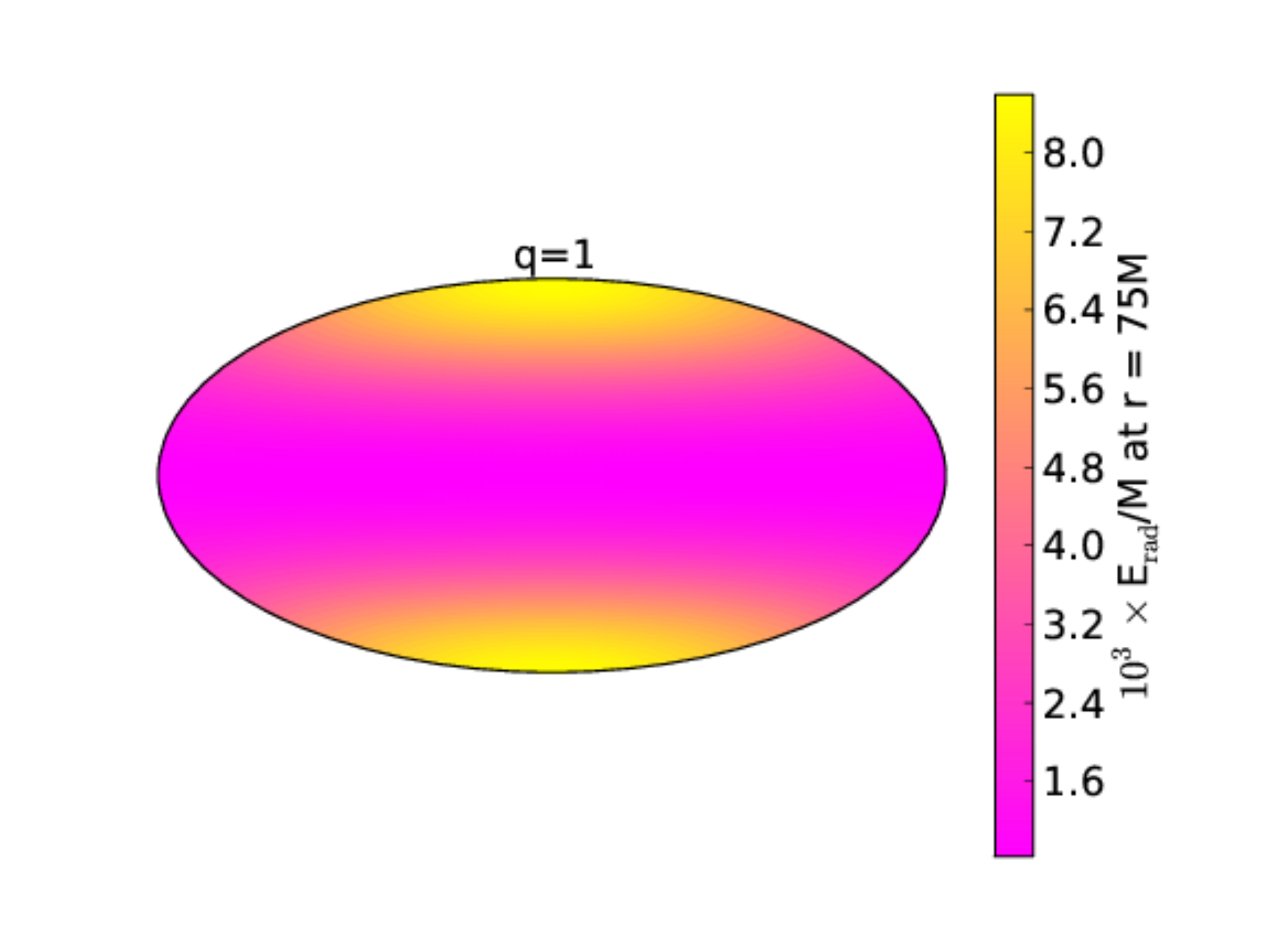}
\includegraphics[width=.5\linewidth]{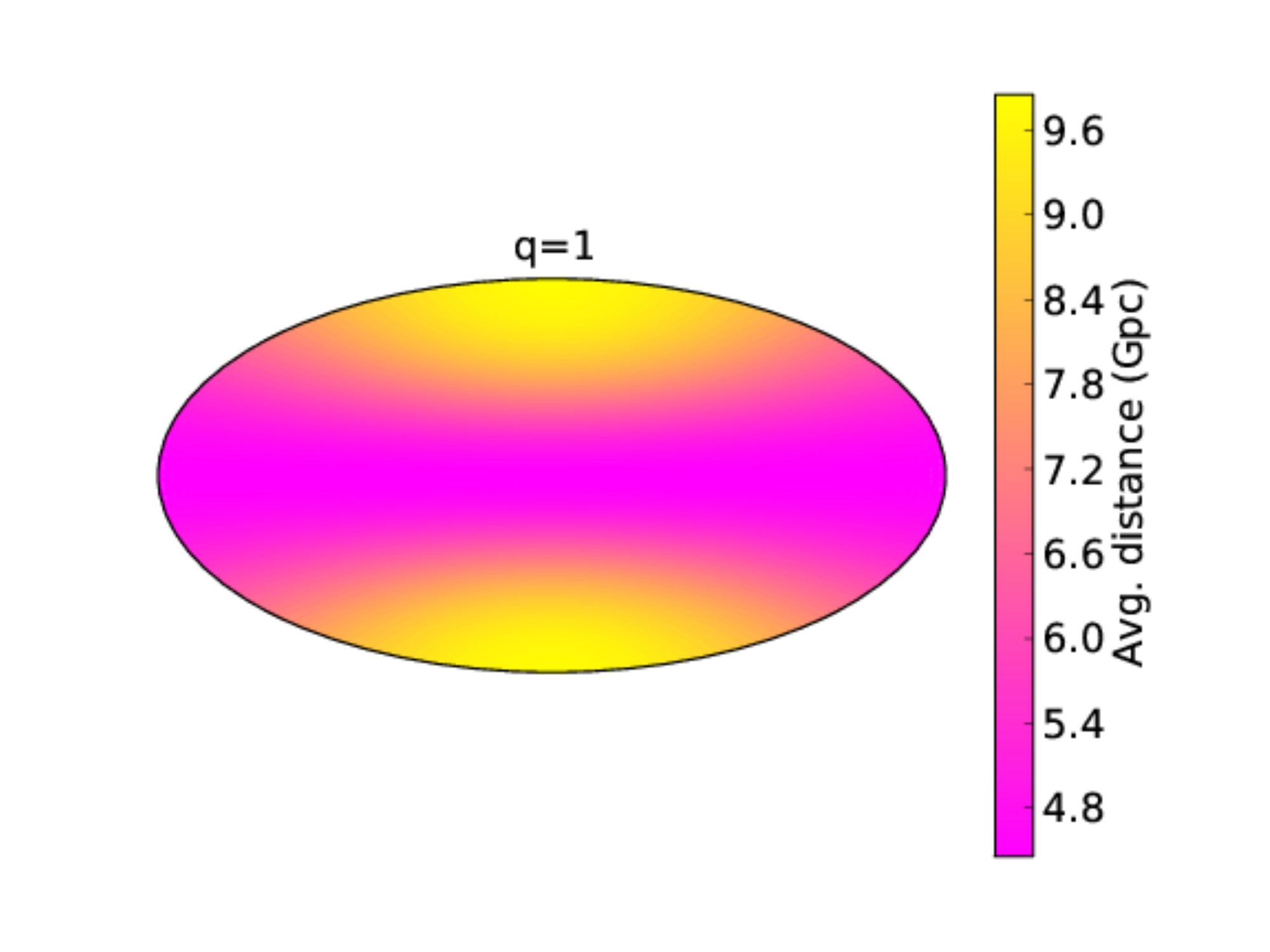}
}
\hbox{
\includegraphics[width=.5\linewidth]{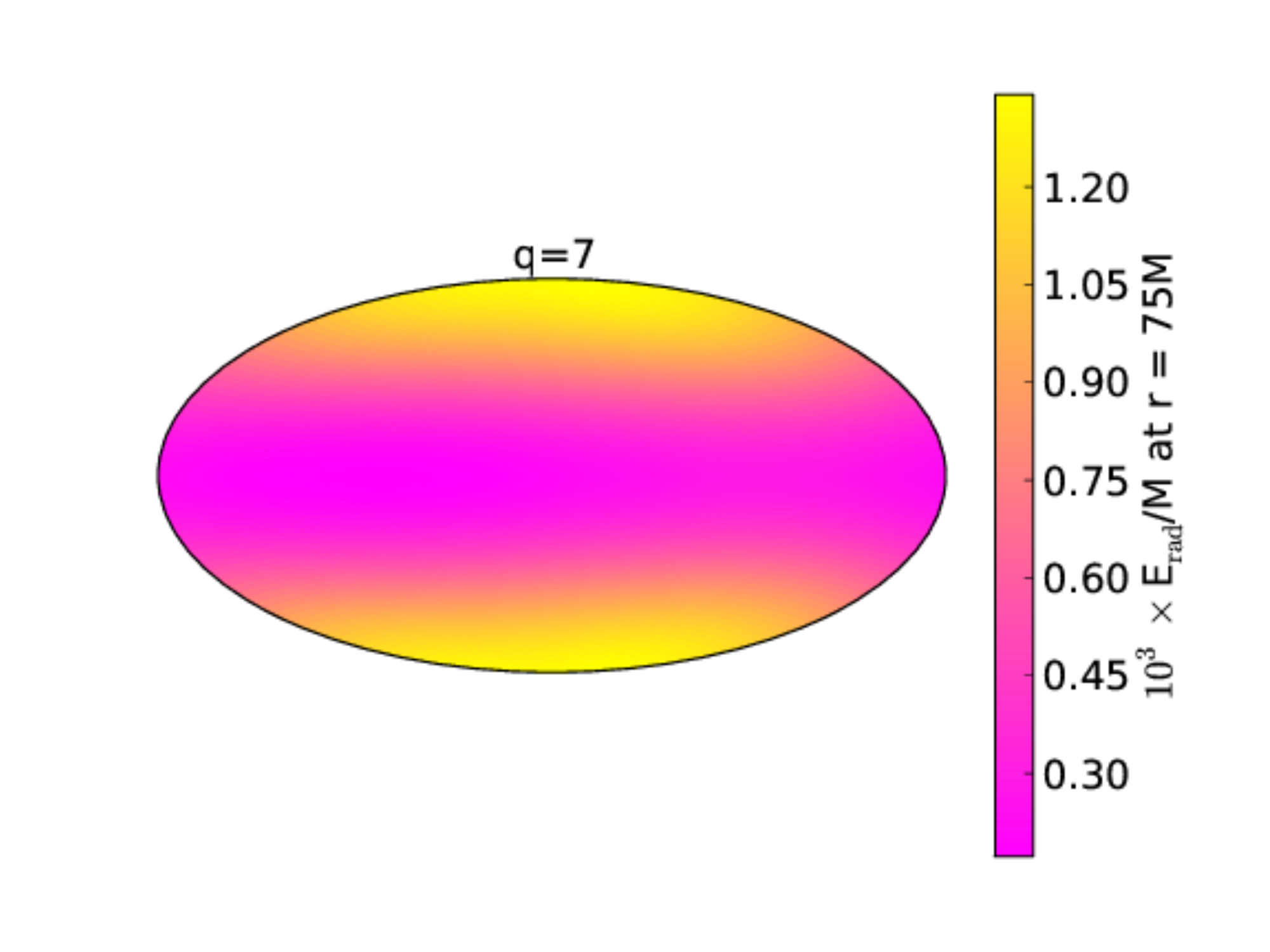}
\includegraphics[width=.5\linewidth]{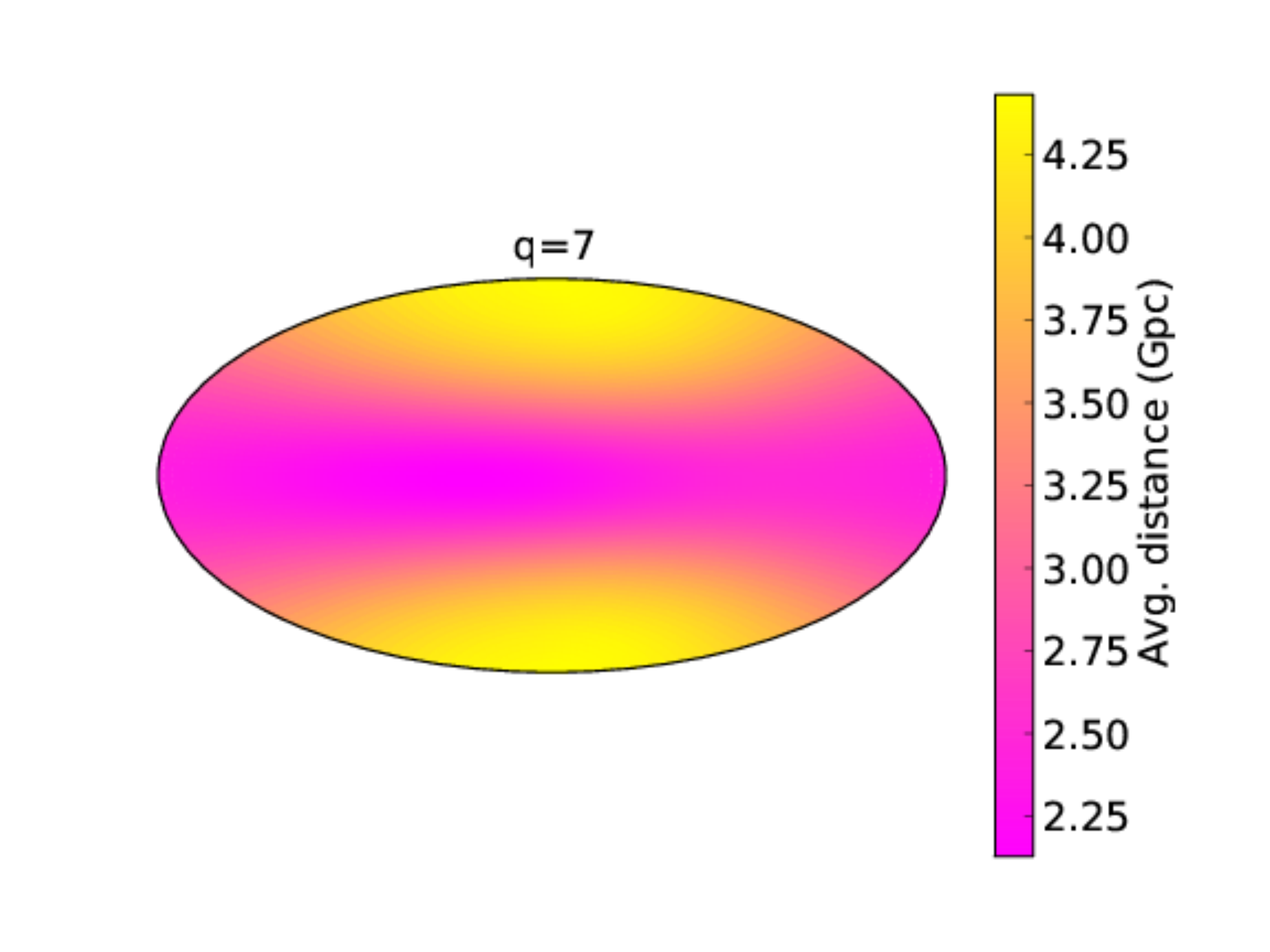}
}
\hbox{
\includegraphics[width=.5\linewidth]{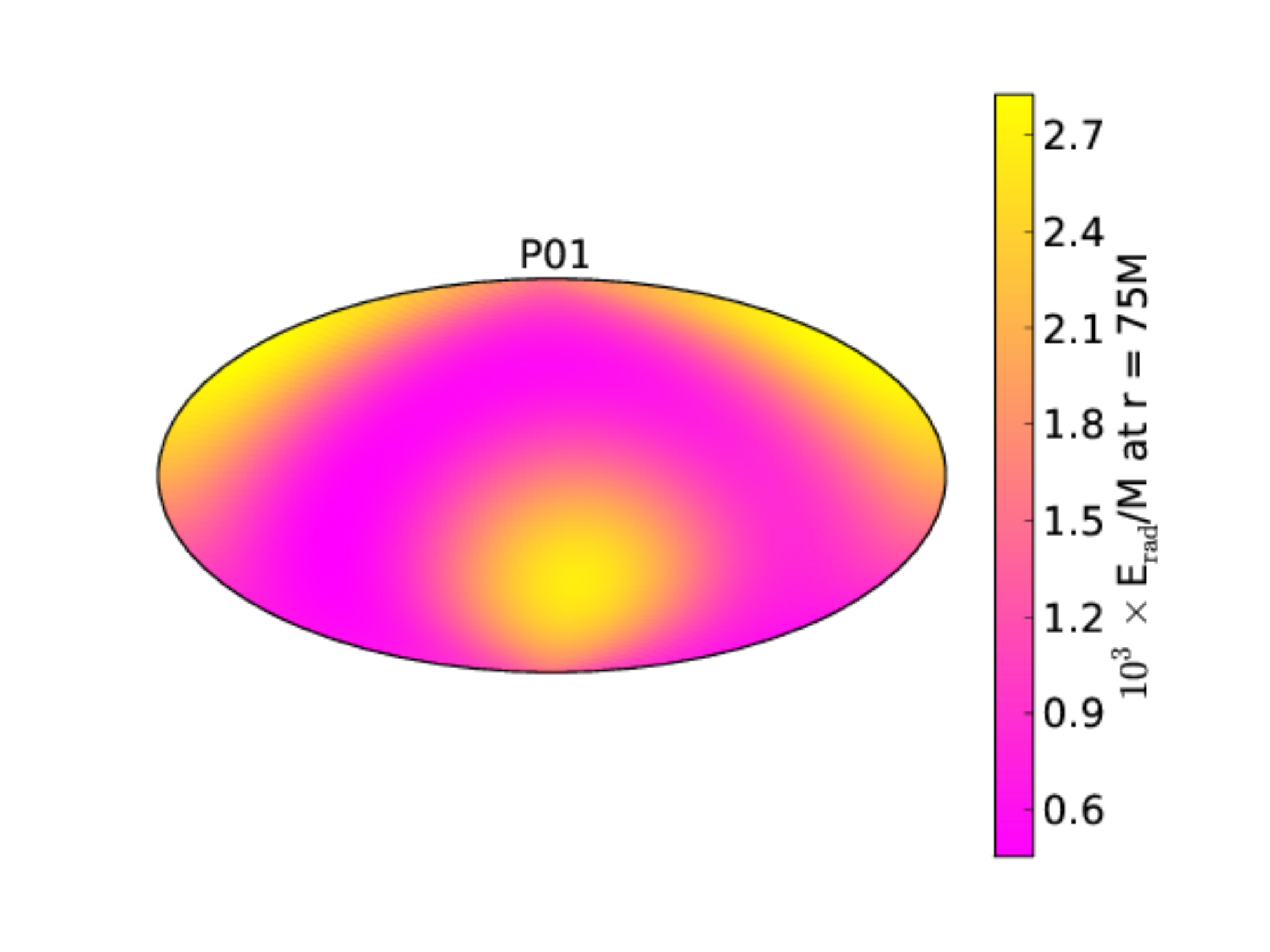}
\includegraphics[width=.5\linewidth]{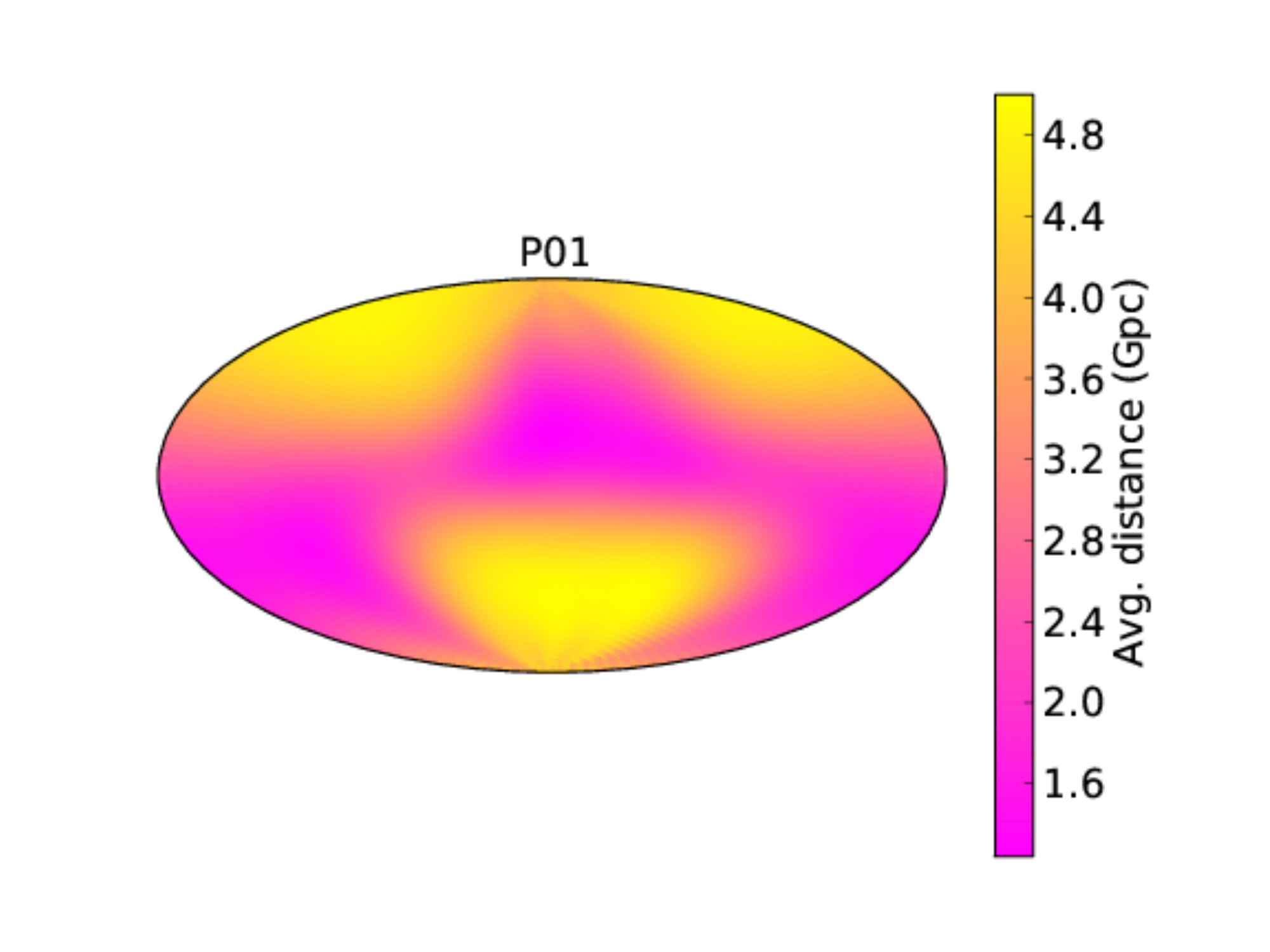}
}
\hbox{
\includegraphics[width=.5\linewidth]{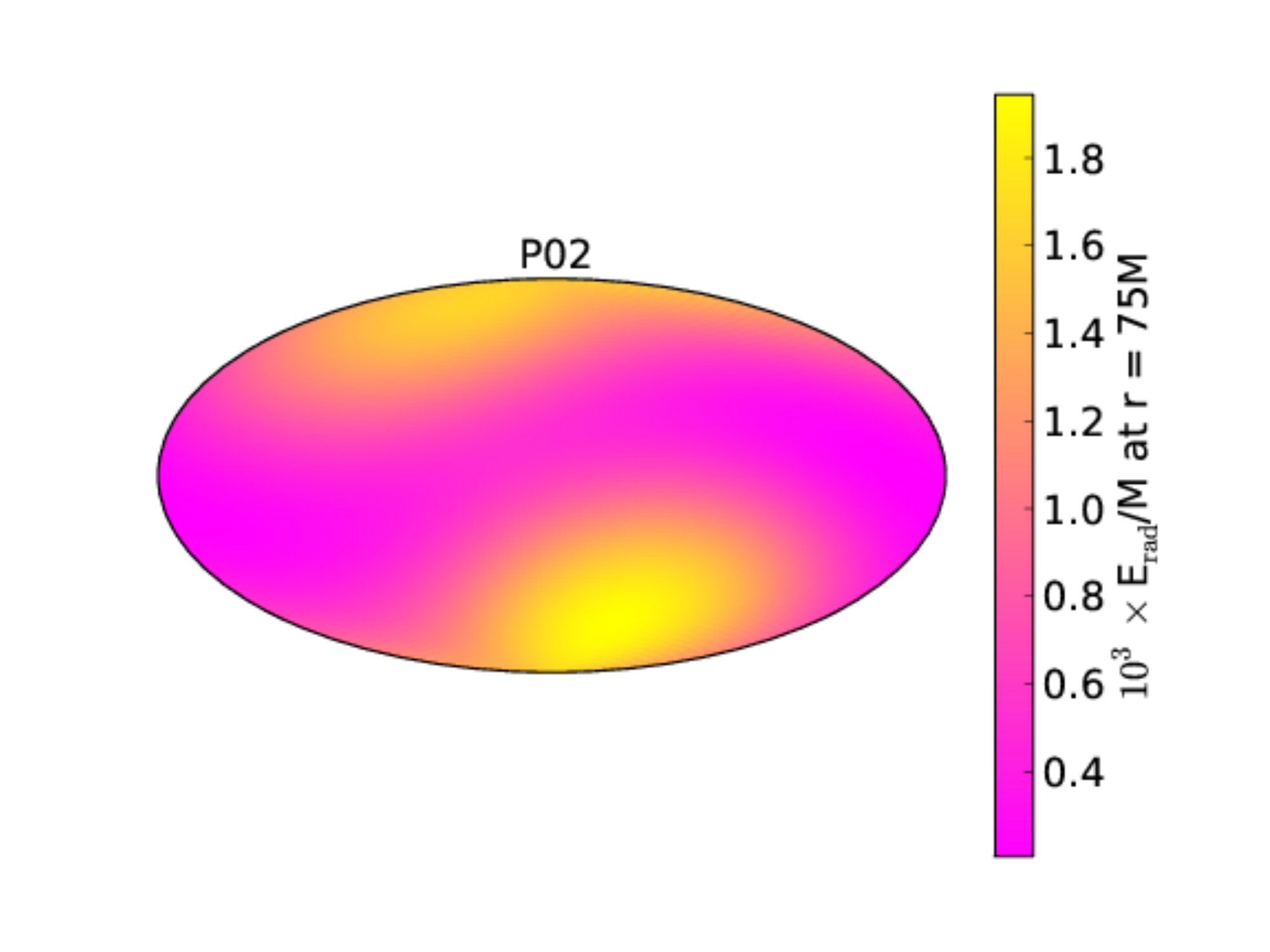}
\includegraphics[width=.5\linewidth]{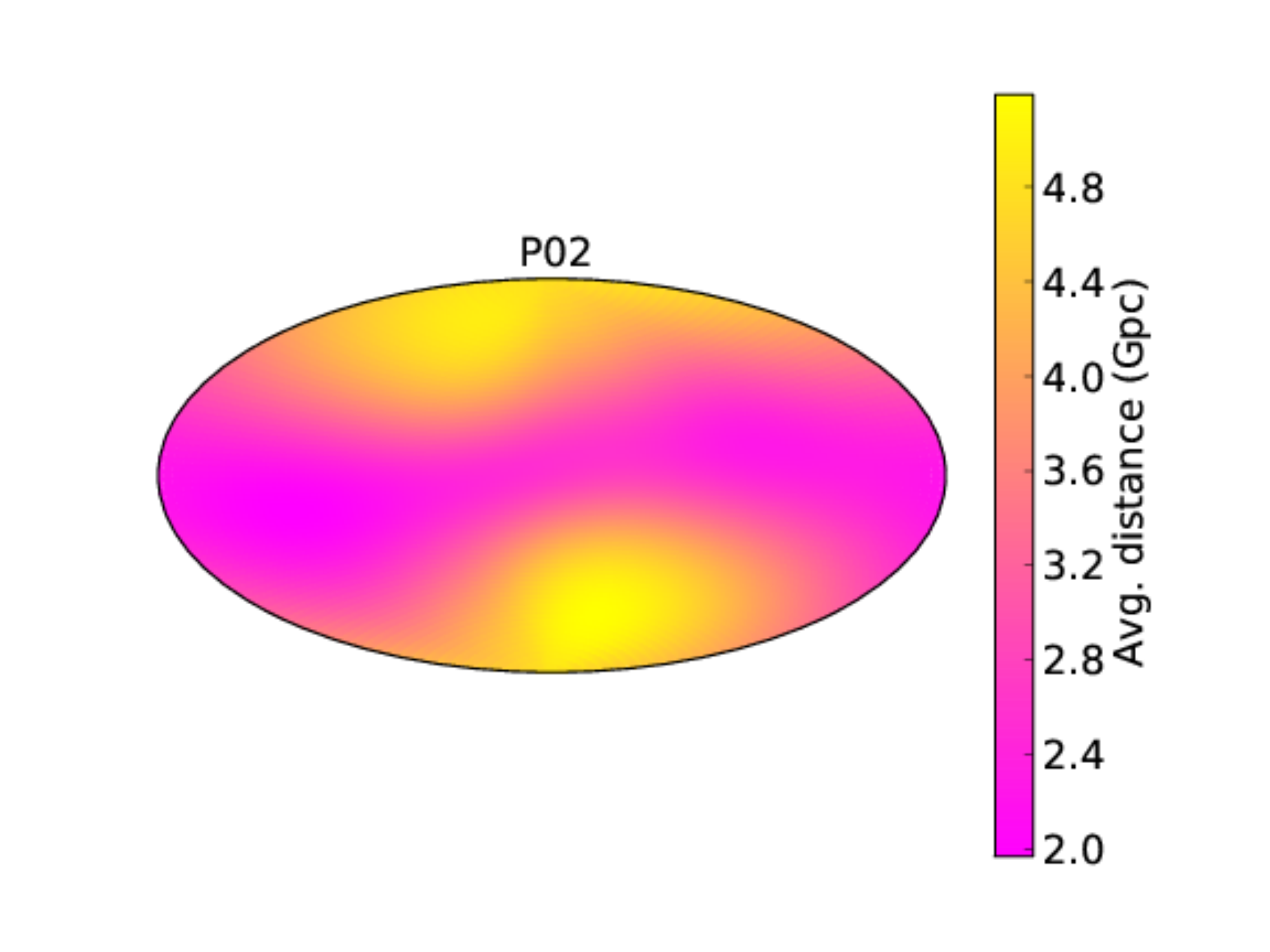}
}
\hbox{
\includegraphics[width=.5\linewidth]{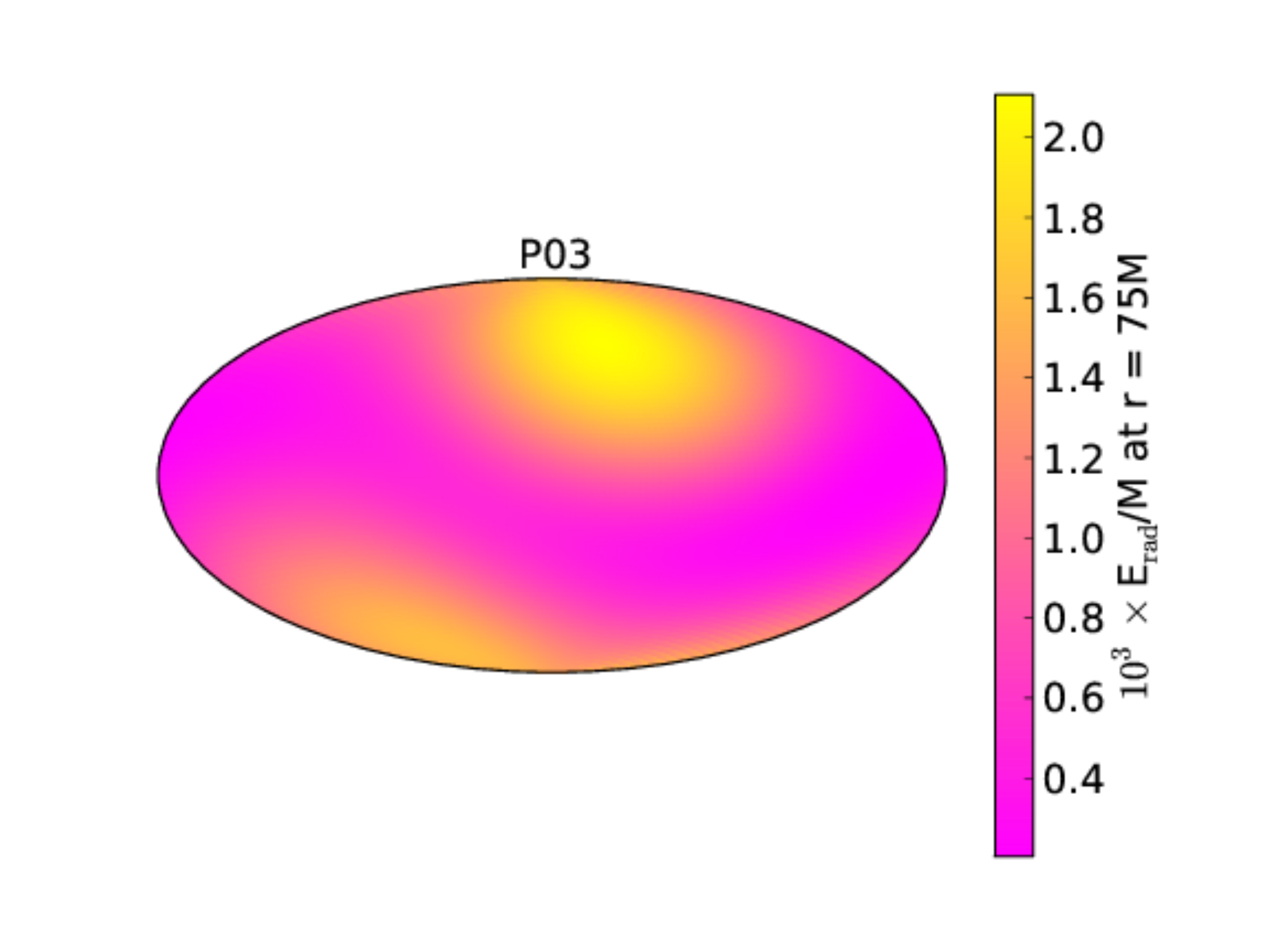}
\includegraphics[width=.5\linewidth]{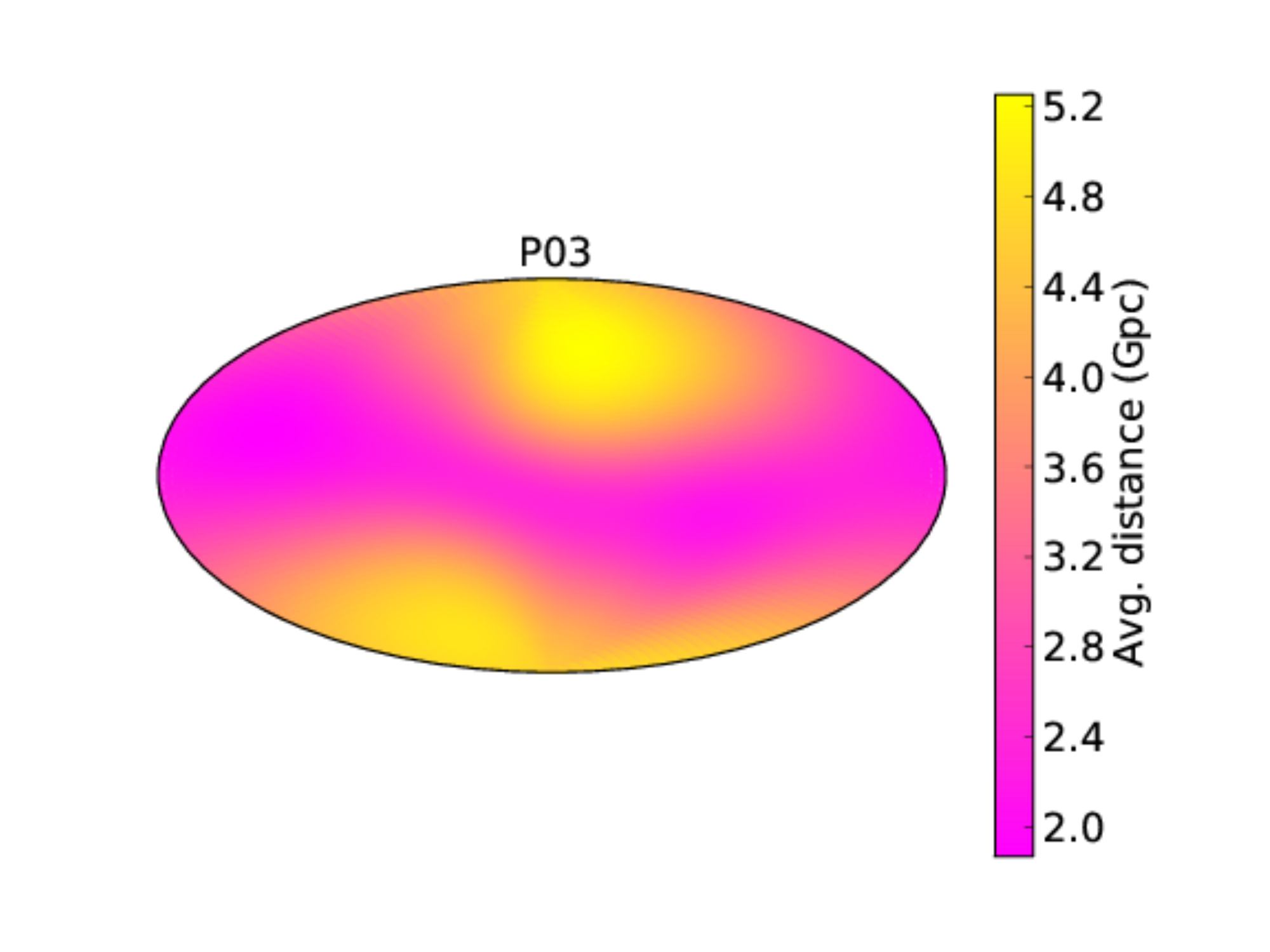}
}
\caption{
Radiated energy and distances to which signals are visible using the
optimal template, in source-centric coordinates $\phi$ horizontally
and $\iota$ vertically.  Top to bottom: $q=1$, $q=7$, and the
precessing P01, P02 and P03 systems.  Note the structure is similar to
the overlaps between the full signal and the $h_{22}$ template,
fig.(\ref{fig:sourceCentricOverlaps}).
}
\label{fig:sourceCentricDistance}
\end{figure}

We start with fig.(\ref{fig:sourceCentricDistance}), which shows the
radiated energy and distances accessible using the $h_{ideal}$
templates, as a function of the source orientation.  As expected, the
range tends to be lowest where the least power is radiated, although
the energy and distance plots are not identical due to weighting by
the noise curve.  The energy, and hence distance, plots have the same
general shape as those corresponding in
fig.(\ref{fig:sourceCentricOverlaps}), indicating that the overlaps
between the signal and $h_{22}$ are lowest at orientations where the
energy and distance reach of the ideal template are also lowest.  This
is due to the fact that the higher modes not only have poor matches
with $(2,2)$, as shown in fig.(\ref{fig:modeOverlaps}), but they also
contain less power, as shown in fig.(\ref{fig:modeAmplitudes}).
Fig.(\ref{fig:sourceCentricDistance}) shows that orientations where
the higher modes dominate have both low matches with $h_{22}$ and
lower ranges.  This indicates that the fractional loss in distance
incurred by using the incorrect template is greatest where the best
possible range is smallest.  

Finally, in order to characterize the performance of different
templates by a single number with physical significance we calculate
the spatial volume to which the search is sensitive.  The distance to
which a signal can be seen depends on all five angles, but from
eqn.(\ref{eq:recievedSignal}) and the comments at the end of
$\S~\ref{sec:matchedfilter}$  the dependence on the detector-centric
angles may be factored out
\begin{equation}
R(\theta,\varphi,\iota,\phi,\psi)
= F_0(\theta,\varphi) R(\iota,\phi,\psi)
\end{equation}

Since there is no preferred orientation we define an {\it average
visibility range}, $R$, by averaging the distances over the
orientation angles $\iota,\varphi,\psi$:
\begin{equation}
\label{eq:snrAvg}
R = \frac{1}{N}
\sum_i^N \frac{\rho(s(\iota_i, \phi_i, \psi_i), h)}{5.5}
\end{equation}
We evaluate this average by choosing random values for $\iota,
\cos(\phi), \psi$ uniform in $(0,2\pi),(-1,1),(0,2\pi)$ respectively.

The average visibility distance as a function of the detector-centric
angles is therefore 
\begin{equation}
R(\theta,\varphi) = R F_0(\theta,\varphi)
\end{equation}
and the volume of the Universe to which a given template is sensitive
is therefore
\begin{eqnarray}
\label{eq:antennaIntegral}
V &=& 
\int_0^{2\pi} d\varphi\,\int_0^{\pi} \sin(\theta) d\theta\,
\int_0^{R(\theta,\varphi)} r^2 dr \nonumber \\
& =&  \frac{1}{3}
\int_0^{2\pi} d\varphi\,\int_0^{\pi} \sin(\theta) d\theta\,
R^3(\theta,\varphi) \nonumber \\
& =&  \frac{R^3}{3}
\int_0^{2\pi} d\varphi\,\int_0^{\pi} \sin(\theta) d\theta\,
F_0^3(\theta,\varphi) \,.
\end{eqnarray}
The remaining integral may be done numerically, yielding a value
$\approx 3.687$.  

\begin{figure}[tb]
\centering
\hbox{
\includegraphics[width=\linewidth]{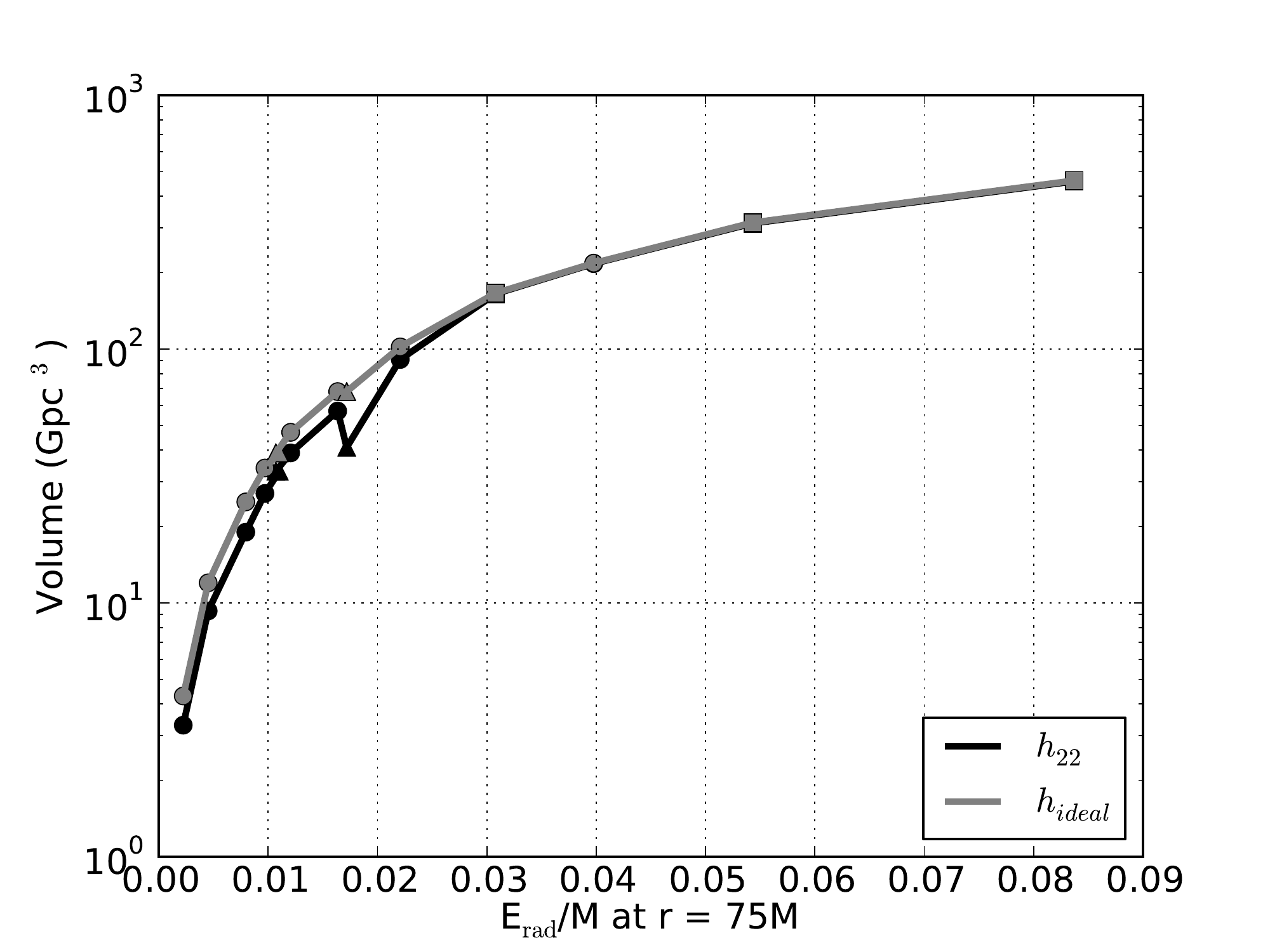}
}
\caption{
Correlation between the total energy radiated from $r=75M$ by the
systems in table~\ref{tab:results} and the accessible volumes using
the $h_{22}$ and $h_{ideal}$ templates.  Circles are non-spinning
systems, squares are spinning but non-precessing systems, and triangles
are precessing systems.  The P02 and P03 systems have close to
identical values of E${}_\textrm{rad}$ and volumes, these points
therefore lie on top of each other.  The $h_{22}$ template gives a
notably smaller fraction of the volume for the P01 system than for
any other, this corresponds directly to the lower overlap noted in 
tab.(\ref{tab:overlapSummaries}).
}
\label{fig:evV}
\end{figure}

The volumes for different waveforms, using the $h_{22}$ and
$h_{ideal}$ templates are summarized in table~\ref{tab:results}.  The
trend is for lower mass ratios and higher aligned spins to correspond
to both larger absolute volumes and smaller relative differences by
including higher modes in the template.  The larger volumes correspond
directly to the increased total energy radiated by such systems, which
is shown in fig.(\ref{fig:evV}).

\begin{figure}[tb]
\centering
\hbox{
\includegraphics[width=\linewidth]{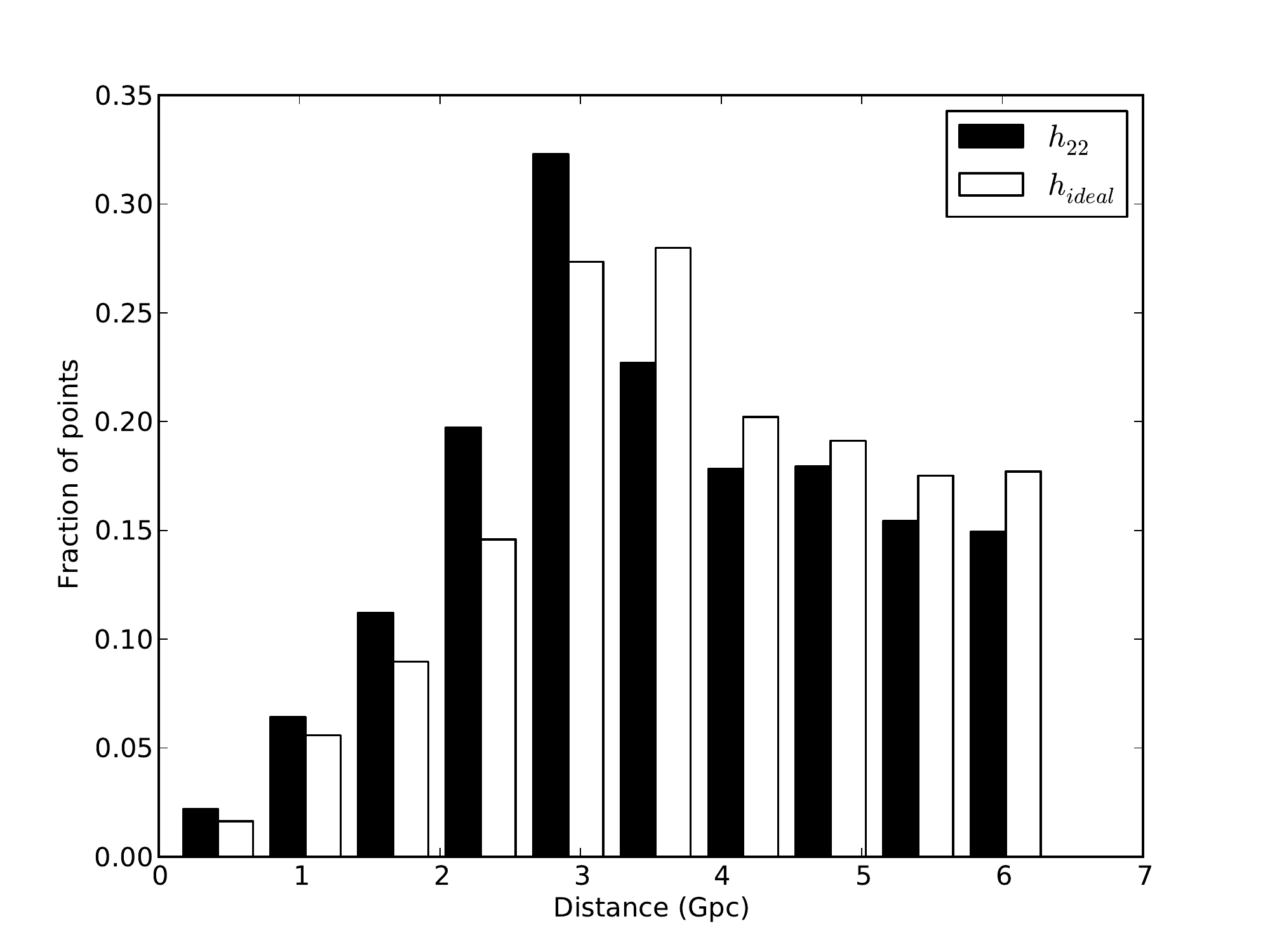}
}
\caption{
Histograms showing the distributions of distances using both
templates for the $q=4$ system.  Using $h_{ideal}$ shifts
points from lower distances to higher, but does not increase
the maximum range.
}
\label{fig:distHist}
\end{figure}

Finally, as another way of quantifying the difference between the
templates, in fig.(\ref{fig:distHist}) we show histograms of the
visibility ranges over the complete set of orientations at $\theta =
\varphi = 0$.  Using $h_{ideal}$ shifts the ranges from lower to
higher values somewhat, but does not increase the maximum distance,
which occurs for face-on systems which are dominated by (2,2).

These results include three precessing $q=4,a=0.6$ systems.  In all
cases the accessible volume is less than that for the $q=4$
non-spinning system.  As might be expected from the non-precessing
cases the volume decreases as the spin becomes anti-aligned with the
angular momentum and less total energy is radiated.  However, at least
for the systems considered here, this dependence becomes smaller than
our uncertainties when the angle between the orbital angular momentum
and the spin of the larger hold exceeds $150$ degrees. 

\begin{table*}
    \begin{tabular}{|r|r|r|r|r|r|r|}
      \hline
      ID & $q$ & $a$ & Volume using           & $R_{avg}$ using& Volume using          & $R_{avg}$ using   \\
          &  &     &   $h_{22}$ (Gpc${}^3$) & $h_{22}$ (Gpc)  & $h_{ideal}$ (Gpc${}^3$) & $h_{ideal}$ (Gpc) \\
      \hline
      H01 &  1 &  0.0 & 217 & 3.3 & 218 & 3.4  \\
      H03 &  3 &  0.0 &  91 & 2.5 & 102 & 2.6  \\
      H04 &  4 &  0.0 &  57 & 2.2 &  68 & 2.3  \\
      H05 &  5 &  0.0 &  39 & 1.9 &  47 & 2.0  \\
      H06 &  6 &  0.0 &  27 & 1.7 &  34 & 1.8  \\
      H07 &  7 &  0.0 &  19 & 1.5 &  25 & 1.6  \\
      H08 & 10 &  0.0 & 9.3 & 1.2 &  12 & 1.3  \\
      H09 & 15 &  0.0 & 3.3 & 0.8 & 4.3 & 0.9  \\
      \hline
      S01 &  1 & -0.4 & 165 & 3.1 & 166 & 3.1  \\
      S02 &  1 &  0.4 & 313 & 3.8 & 315 & 3.8  \\
      S03 &  1 &  0.8 & 458 & 4.3 & 461 & 4.3  \\
      \hline
      P01 &  4 &  0.6 ( $90^{\circ}$) &  41 & 1.9 &  55 & 2.1  \\
      P02 &  4 &  0.6 ($150^{\circ}$) &  33 & 1.7 &  39 & 1.9  \\
      P03 &  4 &  0.6 ($210^{\circ}$) &  33 & 1.8 &  39 & 1.9  \\
      \hline
   \end{tabular}
  \caption[TOADD]{
   \label{tab:results}
   Sensitivity volumes and average distances achievable using both
templates.  ID values correspond to tab.(\ref{tab:NR}) 
Angles following spin magnitude indicate the initial angle of the spin
vector of the larger hole in the $x,z$ plane, such systems exhibit
precession.  Spins not followed by an angle indicate the spins are
(anti) aligned with the orbital angular momentum and the system does
not precess.  Volumes are reduced with increased $q$ and anti-aligned
spins, and increased with align spins due to total power radiated
in-band.  For higher $q$ the use of the ideal template expands the
volume by up to 30\% for the systems considered here, although the 
fractional improvement is greatest for the systems where the volume
accessible with $h_{ideal}$ is smallest.
}
\end{table*}

\subsection{Error analysis}

\begin{figure}[tb]
\centering
\hbox{
\includegraphics[width=.45\linewidth]{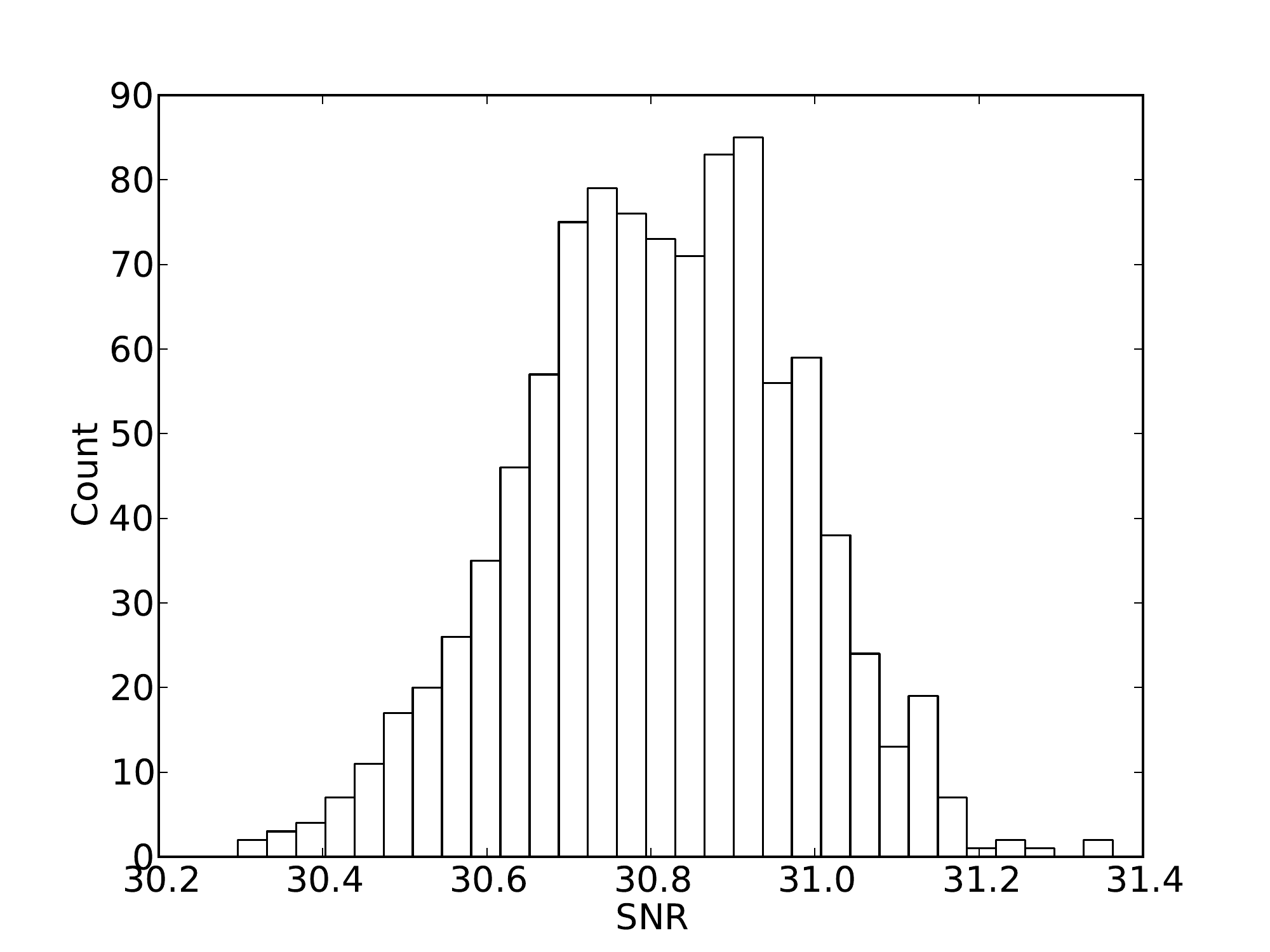}
\includegraphics[width=.45\linewidth]{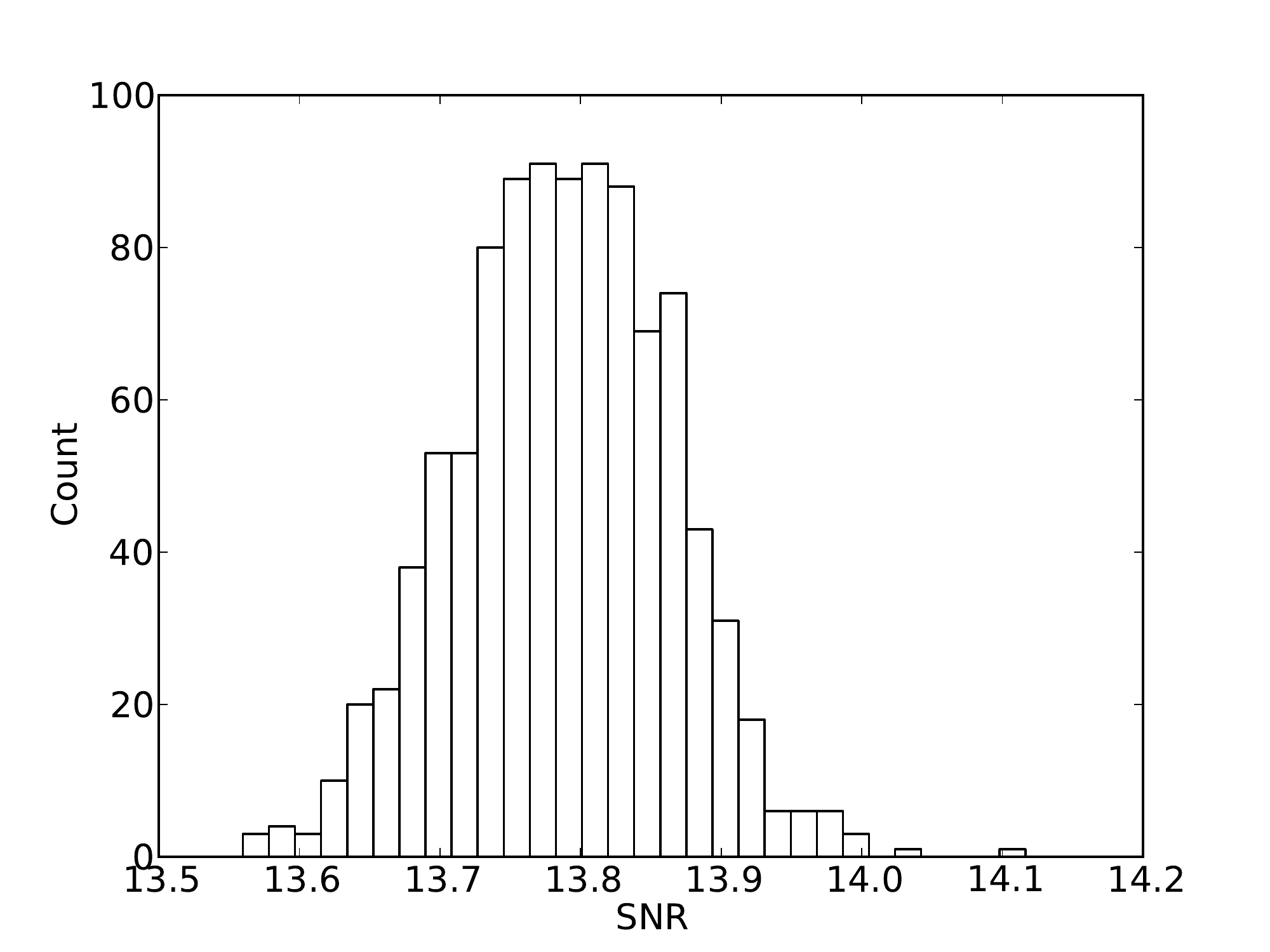}
}
\caption{
Histograms showing variation in distance along the $\theta=\varphi=0$
sky direction for {\bf Left} q=1 and {\bf Right} q=7 systems.  For
$q=1$ the mean is 30.80, corresponding to a distance of 5.6 Gpc, and
the standard deviation is 0.17.  For $q=7$ the mean is 13.79,
corresponding to 2.51 Gpc, and the standard deviation is 0.08.
}
\label{fig:snrAvg}
\end{figure}

Because we choose random values in evaluating the average
eqn.(\ref{eq:snrAvg}) we are able to determine the error in the
results as the standard deviation between several runs.  Due to the
computational expense of complete runs we instead estimate this by
choosing one sky position.  We show the SNR histograms obtained by 900
runs of $\theta=\varphi=\pi/3$ for two waveforms in
fig.(\ref{fig:snrAvg}).  In both cases the error is on order of
$0.5\%$.  Since $V=r^3$ and $r$ has an error $\delta r$, then $ \delta
V = \sqrt{ \left( (dV/dr) \delta r \right)^2 }$.  Here we have $\delta
V / V = 3 \delta r / r $.  The error for the results in
tab.(\ref{tab:results}) is then on order $1.5\%$.  There are also
uncertainties associated with the choice of extraction radius and
resolution.  We show the volumes obtained using the $q=4$ systems and
$h_{ideal}$ template for several value of both parameters in
tab.(\ref{tab:resAndRad}).  The variation is on the order of $1.5\%$,
and our two sources of uncertainty are comparable, and small enough
that they do not effect our conclusions.

\begin{table}
  \begin{center}
    \begin{tabular}{|l||r|r|r|}
      \hline
      \backslashbox{Extraction r.}{Resolution} & M/160 & M/180 & M/200 \\
      \hline
      \hline
      60 M  & 69.59 & 69.62 & 69.64 \\ 
      75 M  & 69.12 & 69.14 & 69.16 \\
      100 M & 68.57 & 68.59 & 68.61 \\
      \hline
    \end{tabular}
  \end{center}
  \caption[TOADD]{
   \label{tab:resAndRad} 
   Volumes obtained using the $q=4$ system and $h_{ideal}$ template
   for various extraction radii and simulation resolutions.  All
   values are in Gpc${}^3$.  All of these runs used the same set of
   points.  There is a general trend downward with decreased
   resolution and increased extraction radius.  The latter effect is
   due to the fact that the late inspiral, merger and ringdown portions 
   of the waveform get smaller as $r\to\infty$.  Although the inspiral portion
   actually increases as $r\to\infty$, since the majority of the power
   radiated is in the last orbits and merger the volume decreases.
   As the variation is small we expect the difference from the true value
   to be small as well.}
\end{table}

\section{Conclusions}
\label{sec:conclusions}
As can be seen from table~\ref{tab:results} there are two conflicting
trends as the mass ratio increases.  As the total radiated energy is
reduced, the volume drops.  Conversely, as the fraction of this energy
is distributed into higher modes the benefit gained by using the ideal
template increases.  The energy radiated, and hence volume, increase
with spin.  Together, these results imply a strong bias towards the
detection of equal-mass, aligned-spin systems when averaged over the sky.  This conclusion is
consistent with~\cite{Reisswig:2009vc,OShaughnessy:2010wa}, while
adding the fact that the inclusion of higher modes is not important
for detecting these systems.  We expect that a search using (2,2)
IMRPhenB aligned-spin templates will perform well, this will be tested
as part of the ongoing NINJA2 project~\cite{Frei2012}.

For non-spinning systems
with $q \gtrsim 3$ and the mildly precessing systems considered here,
the inclusion of higher modes in the template can improve the volume reach of the single detector.
Whether or not this translates to an increase in detection rate
depends on the unknown underlying rates of such systems.  Put another
way, the inclusion of higher modes in templates will allow the
advanced detector network to better measure or bound these unknown
rates.

There are, however, some caveats.  First, we stress that the template
used for the rightmost column of table~\ref{tab:results} exactly
matches the signal, that is, it assumes we exactly know the signal for
which we are looking in advance.  To the extent that matched filtering
is the optimal detection statistic any approximate inclusion of higher
mode information will of necessity do worse.  Furthermore, there are
potential downsides to including higher modes in the templates.  Such
an addition would require increasing the number of
templates.  This entails a corresponding increase in the computational
cost of the search.  In addition, these additional templates may
respond to glitches in the detector, raising the number of
``background'' events and increasing the \snr{} at which a signal
would need to be observed in order to confidently claim a detection.
Concerns such as this lead to changing the mass range in the S6 search
from $35 M_\odot$ to $25 M_\odot$ -- the templates at the higher mass
end produced sufficient numbers of background triggers to impair the
ability to detect lower-mass systems~\cite{LIGO:2012aa}.  It would be
undesirable to allow a search for systems to which the detector
network is comparatively insensitive to impact the ability to detect
equal-mass and aligned-spin systems.  We also note that, at present,
it is not known how to construct a template bank of precessing
signals.  Further studies are needed to determine the right strategy
for detecting both mildly and heavily precessing systems.

We have not yet considered spinning systems with $q > 1$.  Such
simulations are available for spins up to 0.6 and mass ratios up to
7, however, we defer their analysis to future work.  For spins
aligned with the angular momentum the volumes accessible will
certainly be larger than the non-spinning counterparts.  It is
possible that the dependence on higher modes will be preserved in
these cases, leading to a potentially large volume increase by using
templates that include higher modes. 

We have so far considered only a single detector.  Additional
detectors will provide better sky coverage, effectively increasing the
value of the integral in eqn.(\ref{eq:antennaIntegral}).  Furthermore,
as noted in the introduction, detectors oriented differently are
sensitive to different polarizations, it is
therefore conceivable that the inclusion of higher modes in
templates would have more impact on the range of the network as a
whole than on any one detector.  We have also not considered other
aspects of the full search, such as signal-based vetoes.  The effect
of such vetoes is being studied in~\cite{Capano2012}.

One important aspect of gravitational-wave detection we have also not
considered is the fact that the data is filtered against a bank of
templates with different parameters.  For the initial detection it is
acceptable for the signal to be picked up by a template with the wrong
parameters; once the detection has been confirmed more computationally
expensive parameter estimation codes can be run.  While this freedom
can not raise the volume accessible to $h_{ideal}$, as it is already a
perfect match to the signal, it is quite possible that maximization
over a bank of $h_{22}$ templates will lead to larger average SNRs and
hence volumes.  In this case the fractional gain by going to an
approximation of $h_{ideal}$ may be even smaller.  

\begin{figure}[tb]
\centering
\hbox{
\includegraphics[width=.45\linewidth]{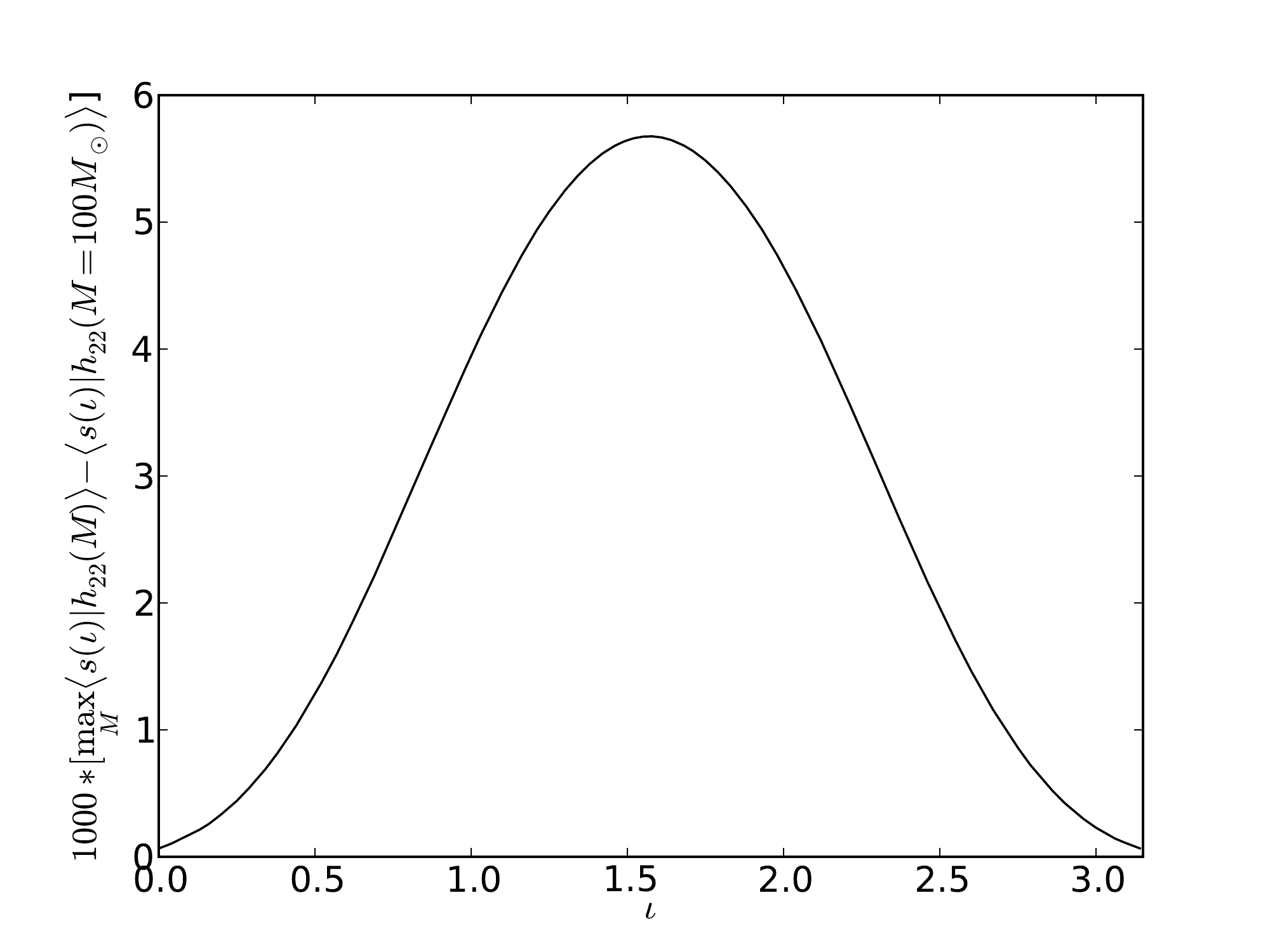}
\includegraphics[width=.45\linewidth]{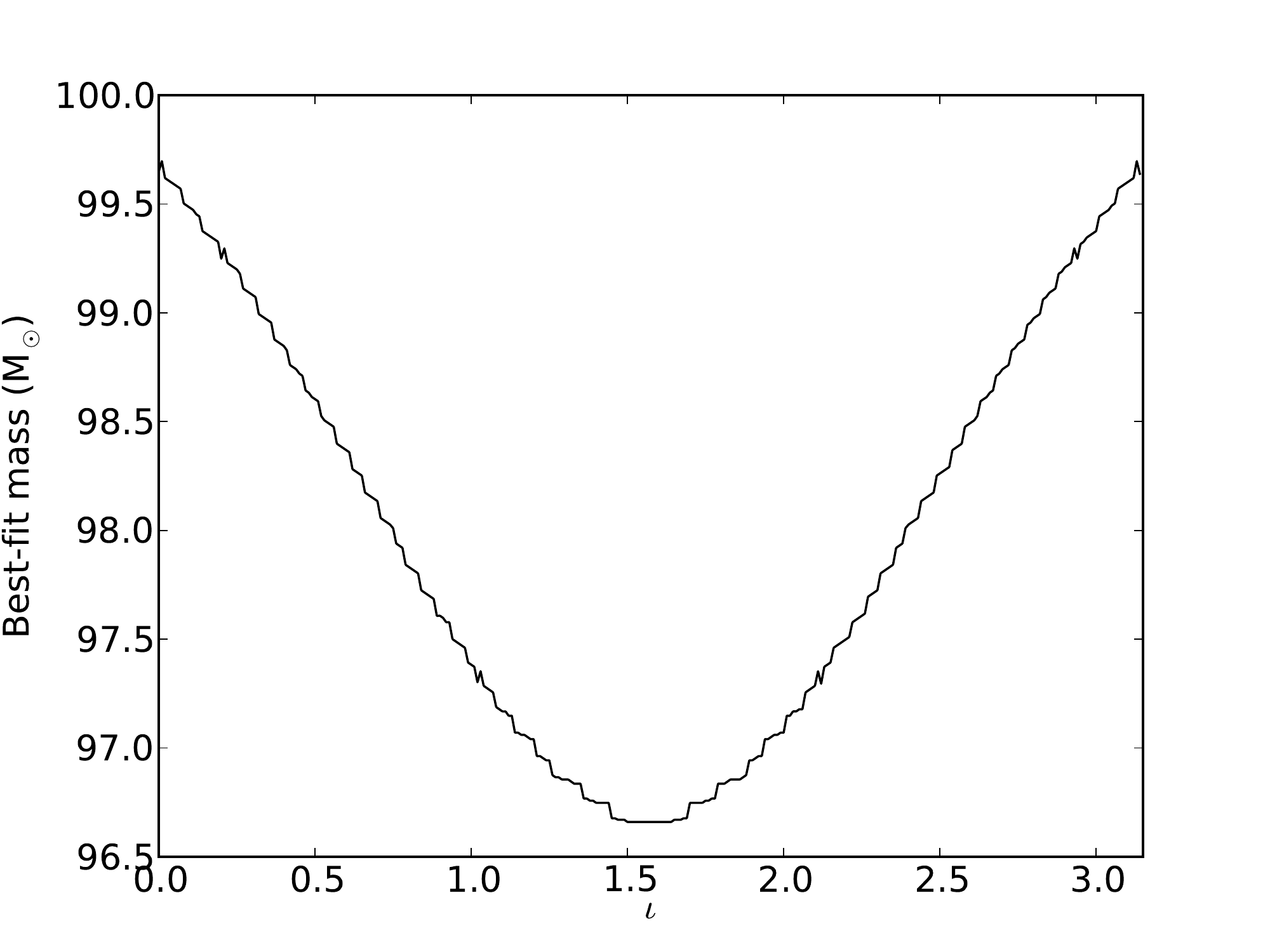}
}
\caption{Effect of higher modes on parameter recovery.
{\bf Left}: the difference in overlap obtained by maximizing
over the mass of the template. {\bf Right}: the value of the mass
which maximizes the overlap.  The largest differences are at
$\iota=\pi/2$, where the system is edge-on and the $(2,\pm 2)$
modes are most suppressed.
}
\label{fig:maxOverM}
\end{figure}

This last point leads to the question of the importance of higher
modes in parameter estimation.  We expect higher modes to be
important here; as a simple example the difference between a signal at
$\iota=\psi=\pi/4$ and one at $\iota=\psi=0$ is entirely encapsulated
in the mode content.  We expect that there are degeneracies between
the orientation parameters and intrinsic parameters, we intend to
investigate this further in subsequent studies.  However we present a
preliminary result in fig.(\ref{fig:maxOverM}), which shows that
$\Overlap{s(\iota,\psi)|h_{22}}$ can be increased by maximizing over
the mass $M$ of the template, at the cost of misestimating the mass.
The increase in overlap is most pronounced at $\iota=\pi/2$, where the
higher modes are most significant.  Correspondingly the mass which
maximizes the overlap deviates the most from the true value at this
point.  This suggests a degeneracy between mass and higher mode
content.  One possible explanation is that the higher modes contain
more power at higher frequencies, as do lower-mass systems.  We will
explore this possibility in our follow-up studies.

\section{Acknowledgments:} Work supported by NSF grants 0914553,
\label{sec:acknowledgements}
0941417, 0903973, 0955825. Computations at Teragrid  TG-PHY120016,
CRA Cygnus cluster and the Syracuse University Gravitation and
Relativity cluster, which is supported by NSF awards PHY-1040231,
PHY-0600953 and PHY-1104371.  This research was supported in part by the National Science Foundation under Grant No. NSF PHY11-25915.  We also would like to thank Stephen
Fairhurst and Duncan Brown  for their  comments. 

\bibliography{nr}

\end{document}